\documentclass[eqsecnum,aps,12pt]{revtex4}
\usepackage{graphicx}
\usepackage{dcolumn}
\usepackage{bm}
\usepackage{color}

\newcommand{\bq}{{\bf q}}
\newcommand{\bv}{{\bf v}}

\newcommand{\br}{{\bf r}}

\newcommand{\bna}{{\bf \nabla}}

\newcommand{\bvp}{{\bf v}_\perp}
\newcommand{\dv}{{\bf \delta v}}
\newcommand{\dvp}{{\bf \delta v}_\perp}
\newcommand{\vpa}{v_\parallel}
\newcommand{\dvpa}{\delta v_\parallel}

\newcommand{\bM}{{\bf M}}
\newcommand{\bff}{{\bf f}}

\newcommand{\bn}{{\bf n}}

\newcommand{\bbr}{{\bf r}}

\newcommand{\hx}{\hat{x}}
\newcommand{\hxp}{\hat{x}_\parallel}

\newcommand{\sep}{ \ \ \ , \ \ \ }

\newcommand{\beq}{\begin{equation}}
\newcommand{\eeq}{\end{equation}}
\newcommand{\beqn}{\begin{eqnarray}}
\newcommand{\eeqn}{\end{eqnarray}}
\newcommand{\pp}{\partial}

\newcommand{\cO}{{\cal O}}

\newcommand{\cred}{\color{red}}

\begin{document}

\begin{center}
{\large {\bf  Why walking is easier than pointing: Hydrodynamics of dry active matter}}
\vskip .75cm
    {\bf  John Toner }
\vskip.5cm
    { Institute of Theoretical Science\\ and
Department of Physics,\\ University of Oregon, Eugene OR
97403-5203, USA}
\end{center}

\section*{Abstract} 
Although human beings have known about the phenomenon of ``flocking"- that is, the coherent movement of large numbers of creatures (flocks of birds, schools of fish, herds of woolly mammoths, etc.)- since prehistoric times, it is only in the last two decades that we have begun to truly understand this phenomenon. In particular, the surprising fact that a very large collection of organisms in two dimensions cannot all {\it point} in the same direction, but can quite easily {\it move} in the same direction, can now be explained. In these lectures, I'll review one of the principle theoretical tools that made this possible: hydrodynamics. My intention is both to 
elucidate flocking- or, to use the specific technical mouthful, "polar ordered dry active fluids"-, and to use flocking as an illustration of how to use the hydrodynamic approach on new and unfamiliar systems. 


\section{Introduction}

Everyone has seen ``flocking", by which I mean the collective, coherent motion of large numbers of
organisms\cite{boids}.  Flocks of
birds and  schools of fish, and herds of wildebeest, are all familiar sights (although the latter possibly only in nature documentaries). Perhaps nowadays it is most commonly seen in the  simulations used for digital cinematic special effects 
 \cite{boids}; these have lead to  the only Oscar ever given for a physics project!
 
 In the past couple of decades,  many synthetic systems of self-propelled particles have been fabricated\cite{Bartolo, Bartolo2} that also exhibit flocking. In addition to providing important experimental realizations of this phenomenon, these experiments make clear that flocking does {\it not} depend on intelligent decision making by the flockers, but, rather, can arise spontaneously from simple short ranged interactions. 
 
 I will hereafter
refer to all such collective motions - flocks, swarms, herds, collections of synthetic self-propelled objects, etc -
as ``flocking''; for convenience, I will also refer  to the ``flockers" as "birds", or, interchangeably, ``boids". 

Note that flocking can occur over an enormous range of length scales:  from
kilometers (herds of wildebeest) to microns (e.g., the
microorganism Dictyostelium discoideum
\cite{dictyo,rappel1, rappel2}).

Remarkably, despite the familiarity and widespread nature of the phenomenon, 
it is only in
the last 24  years that many of the universal features of flocks
have been identified and understood.  It is my goal in these lectures to explain how we've come to understand one particular type of `flocking", namely "polar ordered dry active fluids", which I'll define soon. In the process, I hope to introduce those of you unfamiliar with it to the ``hydrodynamic" approach, which is a powerful technique that can be applied to any large scale collective phenomenon.

To my knowledge, the first physicist to think about flocking - certainly the physicist who kicked off the modern field of active matter- was Thomas Vicsek
\cite{Vicsek}. He was, as far as I know, the first 
to recognize that
flocks fall into the broad category of nonequilibrium dynamical
systems with many degrees of freedom that has, over the past few
decades, been studied using powerful techniques originally developed for
equilibrium condensed matter and statistical physics (e.g., scaling, the
renormalization group, etc).   In particular, Vicsek noted an analogy
between flocking and
ferromagnetism:  the velocity vector of the individual birds is like the
magnetic spin on an iron atom in a ferromagnet.  The usual ``moving
phase'' of a flock, in which all the birds, on average, are moving in the
same direction, is then the analog of the ``ferromagnetic'' phase of
iron, in which all the spins, an average, point in the same direction.
   Another way to say this is that the development of a nonzero mean
center of mass velocity $\left<\bv\right>$ for the flock as a whole
therefore requires spontaneous breaking of a continuous symmetry
(namely, rotational), precisely as the development of a nonzero
magnetization $\vec{M} \equiv <\vec{S}>$ of the spins in a ferromagnet
breaks the continuous \cite{crystalfield}   spin rotational
symmetry of the Heisenberg magnet   \cite{spinspace}.

To make this analogy complete obviously requires that the birds, like the spins in a ferromagnet, live in a rotation invariant environment; that is, that the spins have nothing external that tells them which direction to point, and the birds have nothing external that tells them which way to fly.

To study this phenomenon- the spontaneous breaking of rotation invariance by spontaneous collective motion- which is what I will mean henceforth by the term ``flocking"- Vicsek formulated his now famous algorithm. I will not describe this algorithm in detail-it's probably already been described by others at this school-but will limit myself to noting the features of it that are important for a hydrodynamic theory. These features are: activity, conservation laws, symmetries, short ranged interactions,  and noisiness. To elaborate on these:

\begin{enumerate}

\item  Activity: A large number (a ``flock'') of point particles (``boids''
\cite{boidsterm}) each move
over
time through a space of dimension $d$ ($=2,3$,...),
{\it attempting} at all times to ``follow'' (i.e., move in the same
direction as)
its neighbors. This motion is due to some form of self-propulsion; in Vicsek's algorithm, the rule is that the speed of each creature is constant. Departures from this rule are not important, provided that the boids prefer to be in a state of motion, rather than at rest. This is what is meant by the word ``active" in "polar ordered dry active fluids".

This self propulsion requires an energy source; it also requires that the system be out of equilibrium. Dead birds don't flock!

\item Conservation laws: the underlying model does {\it not} conserve momentum; the total momentum of the flock can change. Indeed, it does so every time a creature turns. We imagine this happening because the creatures move either over  a fixed surface, in two dimension, or through some fixed matrix (e.g., a gel) in three dimensions. This is what is meant by the term ``dry" in "polar ordered dry active fluids". Note that many of the systems you have heard about at this school-e.g., active nematics- are``wet", by which we mean momentum is conserved. Note, incidentally, that real birds (and not only water birds!) are ``wet" in this sense, since the sum of their momentum and the momentum of the air through which they fly is conserved. This changes the dynamics considerably. The problem of wet flocks can still be treated by a hydrodynamic approach\cite{SR}, but the hydrodynamic model is different because of momentum conservation. I will not discuss that case further here. 

There is one conservation law in the Vicsek algorithm, however: the number of birds is conserved. That is, birds are not being born or dying ``on the wing". You laugh, but there are many biological situations- bacteria swarms, and tissue development to name just two - in which this is not a good approximation: bacteria or cells are born and dying on the time scale of the motion. The hydrodynamics of this case is quite interesting\cite{Malthus}, but, again, I won't consider that case in these lectures.

\item Symmetry: the underlying model has complete rotational symmetry:  the flock is
equally likely, a priori, to move in any direction. I will here consider models that do {\it not} have Galilean invariance: that is, they have a preferred Galilean frame. This frame is the one in which the background medium over or through which the boids move is stationary.

\item The interactions are purely short ranged:  in Vicsek's model, each ``boid'' only responds to
its neighbors. In Vicsek's model, these are defined as those ``boids'' within some fixed, finite distance
$R_0$, which is assumed to be independent of $L$, the linear size of the
``flock.'' Hence, in the limit of flock size going to infinity-i.e., the ``thermodynamic limit"- the range of interaction is much smaller than the size of the flock. Variants on this rule-for example, interactions whose strength falls off exponentially with distance- can also be considered  short-ranged.

\item The ``following'' is not perfect: the ``boids'' make errors at all times,
which
are modeled as a stochastic noise.  This noise is assumed to have only short
ranged spatio-temporal correlations. Its role in this problem is very similar to the role of temperature in equilibrium systems: it tends to disorder the flock.
As you'll see, one of the most interesting questions in this problem is whether the ordered state can survive this noise.

\end{enumerate}

In addition to these symmetries of the questions of motion, which
reflect the underlying symmetries of the physical situation under
consideration, it is also necessary to treat correctly the symmetries of
the {\it state} of the system under consideration.  These may be different
from those of the underlying system, precisely because the system
may spontaneously break one or more of the underlying   symmetries of
the equations of motion.  Indeed, this is precisely what happens in the
   ordered state of a ferromagnet:  the underlying rotation
invariance of the system as a whole is broken by the system in its
steady state, in which a unique direction is picked out -- namely,
the direction of
the spontaneous magnetization.

As should be apparent from our earlier discussion, this is also what
happens in a spontaneously moving flock.  Indeed, the symmetry that
is broken -- rotational -- and the manner in which it is broken - namely,
the development of a nonzero expectation value for some vector (the
spin $\vec{S}$ in the ferromagnetic case; the velocity $<\bv>$ in
the flock) are precisely the same in both cases \cite{spinspace}.

The fact that it is a unique {\it vector} that is singled out, rather than merely a unique {\it axis}, is the meaning of the word ``polar" in "polar ordered dry active fluids".

Many different ``phases'' \cite{phases}, in this sense of the word,  of a
system with a given underlying symmetry are possible.  Indeed, I have already
described two such phases of flocks:  the ``ferromagnetic'' or moving flock,
and the ``disordered,'' ``paramagnetic,'' or stationary flock.

In equilibrium statistical mechanics, this is precisely how we
classify different phases of matter:  by the underlying symmetries that
they break.  Crystalline solids, for example, differ from fluids (liquid
and gases) by breaking both translational and orientational
symmetry.  Less familiar to those outside the discipline of soft
condensed matter physics are the host of mesophases known as
liquid crystals, in some of which (e.g., nematics \cite{deGennes}) only
orientational symmetry is broken, while in others, (e.g., smectics
\cite{deGennes}) translational symmetry is only broken in {\it some}
directions, not all.

It seems clear that, at least in principle, every phase known in
condensed matter systems could also be found in flocks.  In these lectures, I'm going to focus one just one phase: the "polar ordered dry active fluid phase", in which rotational symmetry is completely broken by the development of a non-zero average flock speed $\left<{\bf v}\right>$ , but all of the other symmetries of the dynamics (e.g., translation invariance) are preserved.

The first, and to my mind, still the biggest surprise in the entire field of active matter is that a "polar ordered dry active fluid phase" is even possible in two dimensions. 
The reason I (and Vicsek) find this so surprising  is the well-known ``Mermin-Wagner   Theorem''
\cite{MW}   of equilibrium statistical mechanics.
This theorem states
that in a thermal equilibrium model at nonzero temperature with short-ranged
interactions, it is impossible to spontaneously break a continuous
symmetry.  This implies in particular that the equilibrium or
``pointer'' version of Vicsek's algorithm described above, in which
the birds carry a vector $\bv_i$ whose direction is updated
according to Vicsek's algorithm, but in which the birds do not actually
move, can never develop a true long-range ordered state in which all
the $\bv_i$'s point, on average, in the same direction (more
precisely, in which $<\bv> \equiv {\Sigma_i \bv_i \over
N}\neq \vec{0}$) ,  since such a state breaks a continuous symmetry,
namely rotation invariance.

Yet the {\it moving} flock evidently has no difficulty in doing so; as
Vicsek's simulation shows, even two-dimensional flocks with
rotationally invariant dynamics, short-ranged interactions, and
noise-i.e., seemingly all of the ingredients of the Mermin-Wagner
theorem -{\it do} move with a nonzero macroscopic velocity,
which requires  $<\bv> \neq \vec{0}$,
which, in turn, breaks rotation invariance, in seeming violation of the
theorem.

There are  a pair of gedanken experiments that make the very paradoxical and surprising nature of this result more obvious. Both experiments start by putting a million people on a flat, featureless plane in the fog. (This school is clearly {\it not} a good place to perform this experiment: Mont Blanc provides a rather conspicuous ``special direction"!) The featurelessness of the plane, and the fog,  ensure rotation invariance (since they leave the people with no external indication of a preferred direction), while the fog has the further role of ensuring that each person can see only a few of her nearest neighbors.

The first experiment now consists of asking everyone to try to point in the same direction. 
The result is that the people cannot all point in the same direction, no matter how good a job they do at aligning with their nearest neighbors (unless, of course, the alignment is perfect). If they make the slightest errors, those will accumulate over distance, so that, even though a given person may point in roughly the same direction as others not too far away from her, widely separated people will inevitably be pointing in wildly different directions. 

The second gedanken experiment consists of slightly modifying the instructions given to these million folks: now ask them to all  {\it walk} in the same direction.

Amazingly, if this instruction is given to the same people, in the same fog, with the same errors, they {\it can} all walk in the same direction. Moving, apparently, is fundamentally different from 
pointing.

Why? That is the question I will answer in the remainder of these notes.

There is a very simple explanation for this {\it apparent}  ``violation" of the Mermin-Wagner theorem:   one of the essential premises of the Mermin-Wagner theorem does
{\it not} apply to movers:  they are {\it not}   systems
in thermal equilibrium.  The nonequilibrium aspect arises from the motion: you can't move forever in a medium with friction  unless you're alive. And, if you're alive, you're not in thermal equilibrium (that's why we say "cold and dead").

Clearly, motion must be what stabilizes the order in $d=2$:  as described above, the motion is the {\it only}
difference between the pointing and moving gedanken experiments just described.

But {\it how} does motion get around the Mermin-Wagner theorem?
And, more generally, how best to understand the large-scale, long-time
dynamics of a very large, moving flock?

The answer to this second question can be found in the field of
hydrodynamics.

Hydrodynamics is a well-understood subject.  This understanding
does {\it not} come from solving the many ({\it very} many!) body
problem of computing the time-dependent positions $\br_i(t)$ of
the $10^{23}$ constituent molecules of a fluid subject to
intermolecular forces from all of the other $10^{23}$ molecules. Such
an approach is analytically intractable even if one knew what the
intermolecular forces were.  Trying to  compute analytically the
behavior of, e.g., Vicsek's algorithm directly would be the
corresponding, and equally impossible, approach to the flocking
problem.

Instead, the way we understand fluid mechanics is by writing down a
set of continuum equations - the Navier-Stokes equations - for 
continuous, smoothly varying number density $\rho ({\vec r}, t)$ and
velocity $\bv (\br,t)$ fields describing the fluid.

Although we know that
fluids are made out of atoms and molecules, we can
define  ``coarse
-grained'' number density $\rho ({\vec r}, t)$ and
velocity $\bv (\br,t)$ fields by averaging over ``coarse -
graining'' volumes large
compared to the   intermolecular or, in the flocks, ``interbird'' spacing.
On a large scale, even discrete
systems {\it look} continuous, as we all know from close inspection of
newspaper photographs and television images.

In writing down the Navier-Stokes equations, one ``buries one's
ignorance'' \cite{Forster} of the detailed microscopic dynamics of the
fluid in a few phenomenological parameters, namely the mean
density $\rho_0$, the bulk and shear viscosities $\eta_B$ and
$\eta_S$,   the thermal conductivity $\kappa$, the specific heat $c_v$,
and the compressibility $\chi$.  Once these have been
deduced from experiment, (or, occasionally, and at the cost of
immense effort, calculated from a microscopic model), one can then
predict the outcomes of all experiments that probe length scales much
greater than a   spatial coarse-graining scale $\ell _0$ and time scales
$\gg t_0$, a corresponding microscopic time, by solving these
continuum equations, a far simpler task than solving the microscopic
dynamics.

But how do we write down these continuum equations?  The answer
to this question is, in a way, extremely simple:  we write down every {\it
relevant}  term that is not ruled out by the symmetries and conservation
laws of the problem.  In the case of the Navier-Stokes equations, the
symmetries are rotational invariance, space and time translation
invariance, and Galilean invariance (i.e., invariance under a boost to
a reference frame moving at a constant velocity), while the
conservation laws are conservation of particle number, momentum
and energy.

``Relevant,'' in this specification means terms that are important  at
large length scales and long timescales.  In practice, this means a ``gradient
expansion:'' we only keep in the equations of motion terms with the
smallest possible number of space and time derivatives.  For example,
in the Navier-Stokes equations, we keep a viscous term
$\eta_s\nabla^2 \bv$, but not a term $\gamma \nabla^4
\bv$, though the latter is also allowed by symmetry, because the
$\gamma \nabla^4
\bv$ term
involves more spatial derivatives, and hence is smaller,   for slow
spatial variation, than the viscous term we've already got.

Our current theoretical understanding of both dry and wet active matter are based largely on applying the hydrodynamic approach I've just outlined to those systems. The rest of these notes will demonstrate how that is done for the specific case of dry polar active fluids, for which the only symmetry is rotation invariance (``dry" means no momentum conservation, while energy conservation is doesn't apply to any active system, since the very term ``active" implies the existence of an energy source for each particle or flocker).

The remainder of these notes are organized as follows:

in section II, I'll present a highly unorthodox, and extremely handwaving, dynamical ``derivation" of the
Mermin-Wagner theorem, to make it clear that there's something very surprising about the stability of flocks in two dimensions. Then in section III, I'll  review the formulation such a hydrodynamic model for dry active matter (which I will sometimes refer to as ``ferromagnetic flocks")  in
\cite{TT1,TT2,TT3,TT4,NL}). 
In section IV, I'll show how one solves this model, and how that solution implies, among many other results,  that the Mermin-Wagner theorem does {\it not} apply to dry polar active fluids: that is, they can develop long-ranged order, even in two dimensions, even in the presence of noise.
In section V I'll give 
a handwaving argument in the spirit of the derivation of the Mermin-Wagner theorem in section II which explains in physical terms the mechanism that stabilizes long ranged order in two dimensions for flocks.

\section{Dynamical ``Derivation" of the Mermin-Wagner theorem}

You will not, with good reason, see anything like the following derivation in any textbook on statistical mechanics. The usual derivation involves the powerful tools of equilibrium statistical mechanics: Boltzmann weights, Hamiltonians, and the like. Since the Mermin-Wagner theorem was derived for equilibrium systems, for which all these tools are available, it would be completely nuts (to use the technical term) not to take advantage of these tools. 

However, as emphasized by Mike Cates in his lectures at this school (see chapter (\ref{Cates}) of this book), {\it none} of those very powerful tools are available for non-equilibrium systems like flocks. It's therefore useful, I think, to attempt the seemingly crazy stunt of deriving the Mermin-Wagner theorem in a purely dynamical way that can be generalized to non-equilibrium systems. In this way I hope to elucidate exactly what it is about moving that is fundamentally different from pointing, and in particular, how that difference makes long-ranged order literally infinitely more robust in two dimensions in a moving system than a pointing one.

So let's think about those million pointers on the featureless plane in the fog. 
Consider in particular the angle $\theta_i( t)$ between the direction a given pointer labeled by $i$ is pointing at time $t$ and some fixed reference direction. A ``Vicsek-like" algorithm for pointers which try to align with their neighbors is the following updating rule for $\theta_i$:
\begin{eqnarray}
\theta _i (t + 1) = \left< \theta _j (t)\right>_n + \eta_i (t)
\label{sigma rule}
\end{eqnarray}
where the symbol $\left<    \right>_n$ denotes an average over
``neighbors'', which are defined as the set of pointers $j$ satisfying
\begin{eqnarray}
\left|\br_j(t) -  \br_i(t) \right| < R _0 .
\label{circle}
\end{eqnarray}
This allows us to define ``neighbors" even if the pointers are distributed in random positions, rather than on a regular lattice. 

The extra term $\eta_i$ is a random noise that takes 
into account the fact that the pointers will inevitably makes mistakes in aligning with their neighbors.  We'll assume this has zero mean (that is, the pointers are no more likely to err to the left than to the right), and variance $\Delta$, and that it is uncorrelated between pointers $(i,j)$, and between successive time steps. That is,
\begin{eqnarray}
\left< \eta _i (t) \right> = 0
\label{ave eta}
\end{eqnarray}
\begin{eqnarray}
\left< \eta _i (t)\eta _j (t^\prime) \right> = \Delta \delta_{ij}
\delta_{tt^\prime}
\label{ave eta2}
\end{eqnarray}
where 
$\left<    \right>$ without the subscript $n$ denote averages over the
random distribution of the noises $\eta_i(t)$. Here the noise strength $\Delta$ will play the role of temperature, in the sense that larger $\Delta$ will lead to more fluctuations, and hence, presumably, less order. 

The flock evolves through the iteration of this rule.  Note
that the ``neighbors'' of a given pointer do not change on each time
step. To foreshadow where I'm ultimately going here, this is not true for movers, which can change their neighbors due to the differences in the motion of different movers within the flock. This is the fundamental difference between pointers and movers that makes the movers capable of aligning in two dimensions, while the pointers cannot.

But let's not get ahead of ourselves here. Returning to the pointers problem, I note,
as first noted by Vicsek himself, that this model is exactly a simple,
relaxational dynamical model for an equilibrium ferromagnet. That is, if we interpret each unit vector $\bn_i$ that gives the direction the $i$'th pointer is pointing as ain the direction that 
``spins'' carried by each pointer, and update them according to
the above rule, then the model is easily shown to be an equilibrium ferromagnet,
which will relax to the Boltzmann distribution for an equilibrium
Heisenberg model (albeit with the ``spins'' living not on a
periodic lattice , as they usually do in most models and in real
ferromagnets, but, rather, on a random set of points). In the absence of noise (i.e., for $\Delta=0$), this algorithm will, unsurprisingly, lead to a ``ferromagnetic" state, characterized by
a non-zero ``magnetization": 
\beq
\bM\equiv\left< \bn \right> \equiv {\sum_{i=1}
\bn_i \over N } \quad ,
\label{Mag}
\eeq
where in this expression the $<>$ mean an average over all the pointers. I'll assume throughout these notes that this average is equal to an average over the noise; in equilibrium physics, this is sometimes called the assumption of ``ergodicity".

At zero noise, we would expect  to, and do,  eventually reach a state in which $ |\bM|=1$; i.e., perfect alignment of all the pointers. The big question is: what happens when there is noise ( i.e., when $\Delta\ne0$)?  

To answer this, begin by noting that the dynamical rule (\ref{sigma rule}) is actually a disguised version of a noisy diffusion equation. To see this, recall that one of the numerical algorithms for solving Laplace's equation $\nabla^2\theta=0$ is to replace the value of the field $\theta$ at each point  at each point with the average of its neighbors. Thus, in the absence of noise, the dynamics (\ref{sigma rule}) will eventually relax the field $\theta$ to a state in which $\nabla^2\theta=0$, which implies that the rate of change of $\theta$ (again, in the absence of noise) is itself proportional to $\nabla^2\theta$ (since it vanishes when $\nabla^2\theta=0$). Indeed, one can very simply derive this result as follows:

Consider  for simplicity (although it is not necessary) a two dimensional collection of pointers arranged on a square grid of lattice constant $a$. The pointer at position $\br_i=(x_i,y_i)$, where my $x$ and $y$-axes are aligned with the square grid, has four neighbors, one to its right at $\br_{1}=(x_i+a,y_i)$, a second to its left at $\br_{2}=(x_i-a,y_i)$, a third above at $\br_{3}=(x_i,y_i+a)$, and the fourth below at $\br_{4}=(x_i,y_i-a)$. Thus, the dynamical rule (\ref{sigma rule}) can be rewritten:
\begin{eqnarray}
&&\theta _i (x_i, y_i, t + 1)-\theta _i (x_i, y_i, t) =\nonumber\\&&{1\over4} \large[ \theta(x_i+a,y_i, t)+\theta(x_i-a,y_i, t)+ \theta(x_i,y_i+a, t)+\theta(x_i,y_i-a, t)\large] -\theta _i (x_i, y_i, t)+ \eta_i (t) \,,\nonumber\\
\label{dif1}
\end{eqnarray}
where I have subtracted the value $\theta_i(t)$ of $\theta_i$ on the last time step from both sides, so as to make the right hand side look like a discrete representation of a time derivative. In the process, I have made the left hand side into a discrete version of the Laplacian. To see this, just reorganize the right hand side  as follows:
\begin{eqnarray}
\theta _i (x_i, y_i, t + 1)-\theta _i (x_i, y_i, t) &=&{1\over4} \large[ \{\theta(x_i+a,y_i, t)-2\theta _i (x_i, y_i, t)+\theta(x_i-a,y_i, t)\}\nonumber\\&&+ \{\theta(x_i,y_i+a, t)-2\theta _i (x_i, y_i, t))+\theta(x_i,y_i-a, t)\}\large] + \eta_i (t) \,.\nonumber\\
\label{dif2}
\end{eqnarray}
Now note that, just as the left-hand side can be approximated as the time derivative of $\theta$ if $\theta$ varies slowly in time - that is, we can write $\theta _i (x_i, y_i, t + 1)-\theta _i (x_i, y_i, t)\approx\partial_t\theta$, the term in the first parenthesis on the right hand side can  be approximated by the second derivative of $\theta$ with respect to $x$:  $\theta(x_i+a,y_i, t)-2\theta _i (x_i, y_i, t)+\theta(x_i-a,y_i, t)\approx a^2\partial_x^2\theta $, provided that $\theta$ varies slowly with popsition. Likewise, the second term can be approximated  by the second derivative of $\theta$ with respect to $y$:  $\theta(x_i,y_i+a, t)-2\theta _i (x_i, y_i, t))+\theta(x_i,y_i-a, t)\approx a^2\partial_y^2\theta$.
 Hence, equation (\ref{dif2}) can be approximated as 
 \begin{eqnarray}
\partial_t\theta &=&{a^2\over4} \large[ \partial_x^2\theta+\partial_y^2\theta\large] + \eta=D\nabla^2\theta+\eta \,,\nonumber\\
\label{diffusion}
\end{eqnarray}
where I've defined the ``diffusion constant" $D\equiv{a^2\over4}$.

Note that I could also have derived this result purely on symmetry grounds: $\partial_t \theta$ must be a scalar made out of $\theta$ itself and its derivatives. By rotation invariance,  it must vanish if $\theta$ is spatially uniform. By the isotropy of space,   it must be an isotropic operator. The only thing you can make that does this to second order in gradients of 
$\theta$ is $\nabla^2\theta$.

So what  are the consequences of the fact that $\theta$ obeys a diffusion equation? There are two that are important for our discussion:

\noindent 1) $\theta$ is slow, and  

\noindent 2) $\theta$ is conserved (in the absence of noise).

To be more precise about point 1), the form of the diffusion equation implies that an initially localized departure of $\theta(\br, t=0)$ from spatial homogeneity  spreads very slowly. One can read this off by power counting from the form of the diffusion equation: a time derivative of $\theta$ can be estimated as roughly $\theta$ over a time $t$, while the Laplacian of $\theta$ can be estimated as $\theta$ divided by a distance $r$ squared. Equating these gives
\beq
r^2\propto t  \,,
\label{difscale1} 
\eeq
or, equivalently, 
\beq
r\propto \sqrt{t}  \,.
\label{difscale2} 
\eeq
The exact solution (in the absence of noise) of the diffusion equation in $d$-spatial dimensions for an initially localized $\theta$, which is
\beq
\theta(\br,t)=A\exp\left(-{r^2\over4Dt}\right)/(4\pi Dt)^{ d\over2} \,,
\label{difsol} 
\eeq
clearly obeys this scaling law. 

This is very slow; indeed, anything moving at any constant speed, however small, will eventually outrun diffusive spreading, since $t\gg\sqrt{t}$ as $t\to\infty$. This is why you stir your coffee after adding milk to it: even a slow stir leads to far faster mixing than diffusion. We'll see later that the reason flocks can order in two dimensions is essentially that, by their motion, they stir themselves.

Turning now to point 2), we can see that $\theta$ is conserved in the absence of noise by setting $\eta=0$ and integrating both sides of the diffusion equation (\ref{diffusion}) over all $\br$. This gives
\beq
{d\over dt}\int d^dr \, \theta(\br, t)= D \int d^dr \, \nabla^2\theta(\br, t) =D\int d^dr \, {\bf \nabla}\cdot {\bf \nabla}\theta(\br, t) = D\int_S d{\bf A}\cdot \, {\bf \nabla}\theta(\br, t) \,,
\label{thetacons}
\eeq
where in the last equality I've used the divergence theorem, with $S$ being the surface bounding my volume of integration. This shows that the integral of $\theta$ over any region of space can only change if there is a current (proportional to $\nabla\theta$) through the surface of that region.  Basically, $\theta$ acts like the milk in your  coffee: the total quantity of it is conserved; diffusion can only redistribute it in space. And it can only do {\it that} very slowly; i.e., the spatial spread $r(t)$ after a time $t$ only grows like $\sqrt{t}$.

\begin{figure}
 \includegraphics[width=1.0\textwidth]{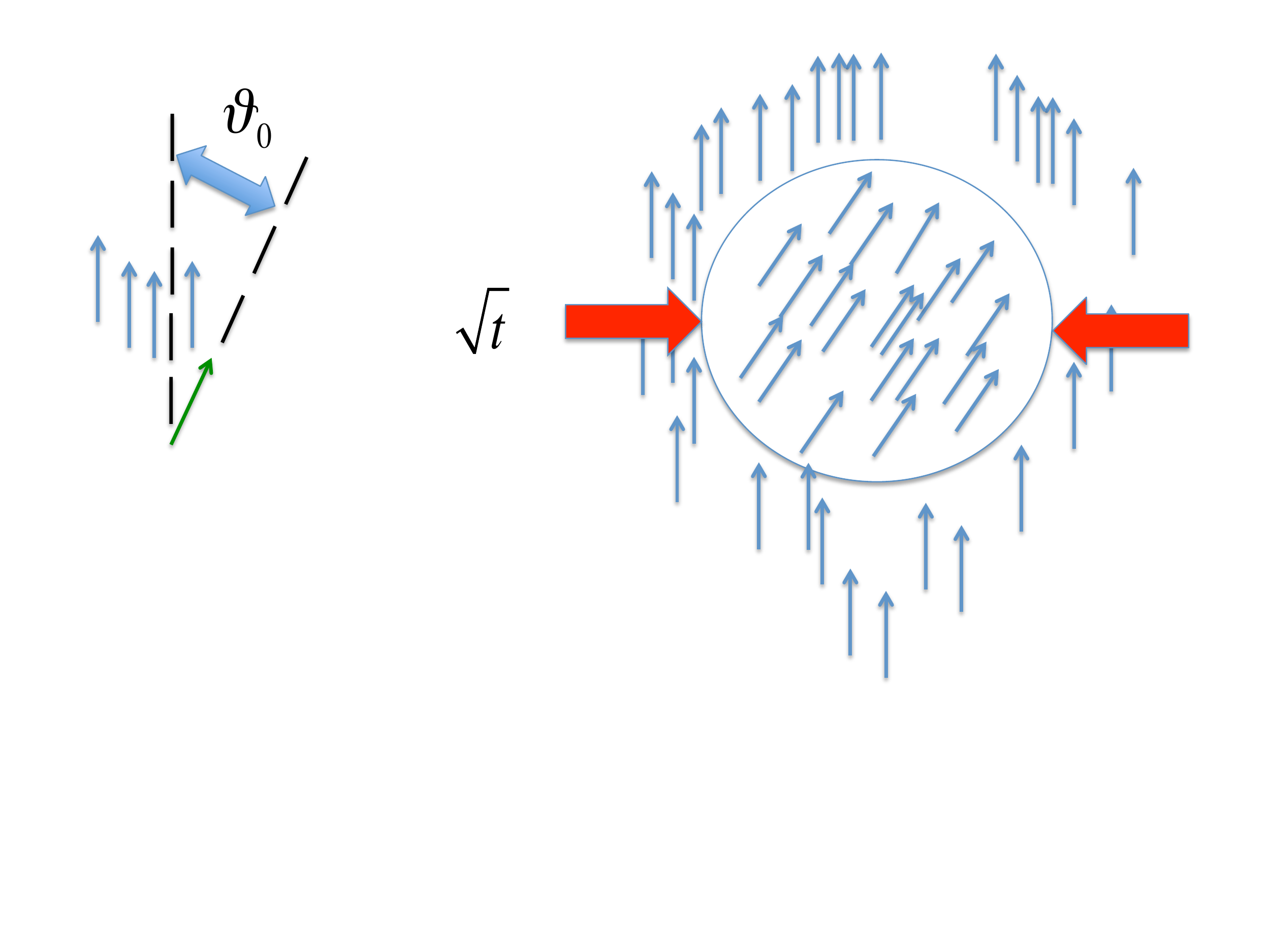}
 \caption{\label{oneerrorfig}Evolution of a single error in the pointer problem. The original error $\theta_0$ gets shared evenly after a time $t$ among all of the pointers within a distance $\propto\sqrt{t}$ of the original pointer.}
\end{figure}

The consequence of these two observations is that fluctuations decay very slowly in the pointer system. To illustrate this dramatically, consider a pointer system with no noise, and with an almost perfectly ordered initial condition: only the pointer at the center is pointing in a different direction from any of the others, and he is pointing at an angle $\theta_0$ to the left of the direction the others are pointing. (See figure (\ref{oneerrorfig})).  Defining $\theta=0$ as the direction the others are pointing, we have $\int d^dr \, \theta(\br, t=0)=\theta_0$.

What will this collection look like after time $t$, if there is no noise? Well, by point 1) above, the initial error will now be spread out over all the pointers within a distance $r(t)\propto\sqrt{t}$. By point 2), the sum of the deviations of all of these pointers (including the original error making pointer)  from the original direction of most of them must still be $\theta_0$, since $\theta$ is conserved. So the original fluctuation of $\theta_0$ must now be distributed over all of those pointers within that distance $r(t)\propto\sqrt{t}$. Hence, we  can crudely estimate the angular deviation after a time $t$ by assuming (as proves to be the case) that this initial error is spread roughly uniformly over all $N(t)$ of the pointers within this distance 
$r(t)$. That number $N(t)$ is easy to estimate; it's just the density times the volume (or hypervolume, if we're considering $d\ne3$) of the region of radius $r(t)\propto\sqrt{t}$. Assuming the density is roughly constant, at least over  a sufficiently large region (as indeed it is for a random set of points; fluctuations in the density of a random set of points over a volume $V$ scale as 
${1\over\sqrt{V}}\to0$ for $V\to\infty$), it's clear that 

\beq
N(t)\propto [r(t)]^d\propto t^{d/2} \,,
\label{Ndif}
\eeq
where I've used $r(t)\propto\sqrt{t}$.
 
Since the original total error of $\theta_0$ is now divided among all $N(t)$ of these pointers, the typical fluctuation $\theta(t)$ of each of them, including the original error-maker, is now  
\beq
\theta(t)\approx{\theta_0\over N(t)}\propto {\theta_0\over t^{d/2}} \,.
\label{one error}
\eeq

I want to call your attention to two things about this result:

\noindent 1) the decay is extremely slow; specifically, it is a power law in time. Hence, it is asymptotically slower than {\it any} exponential decay. This is a consequence of the conservation law for $\theta$, which is in turn a consequence of the underlying rotation invariance. This means $\theta$ is a {\it Goldstone mode} of the system, a concept that may already be familiar to some of you. I'll discuss this more in the next section.

\noindent 2) The power law of this decay is dimension-dependent, with slower decay in lower dimensions. This is a general and recurring theme in statistical and condensed matter physics: fluctuations decay more slowly, and, hence, are more important, in lower dimensions. Ultimately, this is why the Mermin-Wagner theorem applies to low-dimensional systems-specifically, $d\le2$-, but not higher dimensional ones.

With this result (\ref{one error}) for a single initial error in hand, let's now go back and consider our original model with noise. Now the situation is even worse: while any initial errors are very slowly decaying according to (\ref{one error}), more errors are constantly being made. The question now becomes, can the slow decay of (\ref{one error}) keep up with the accumulation of new errors? Given the dimension dependence of (\ref{one error}), you won't be surprised to learn that the answer to this question is also dimension dependent: the errors can be kept under control for spatial dimensions $d>2$, but not for $d\le2$. This is the Mermin-Wagner theorem\cite{MW}.

To see this, consider the ``blob" of pointers that can have exchanged information diffusively with some central pointer after a time $t$. As noted earlier, this blob will have radius $r(t)\propto\sqrt{t}$, or, equivalently, given a radius $r$ of the blob,m the time required for all parts of that blob to be able to communicate with each other is
\beq
t\propto r^2 \,.
\label{tdif}
\eeq
This blob will contain will contain $N(t)\propto r^d$ pointers. How many errors will these pointers collectively have made? Well, each of them will have made
\beq
\#\,{\rm of\, errors/pointer}\propto t\propto r^2 \,;
\label{number of errors/spin}
\eeq
hence, the full collection of $N(t)\propto r^d$ of them will have made
\beq
{\rm total}\, \#\, {\rm of\, errors}\propto N(t)t\propto r^{d+2} \,.
\label{number of errors}
\eeq
Since the sum of a number of independent random variables with zero mean is proportional to the square root of that number, we have
\beq
\sqrt{<\theta^2>}\approx{\sqrt{{\rm total}\, \# \,{\rm of\, errors}}\over N(t)}\propto {r^{d+2\over2}\over r^d}\propto r^{1-d/2} \,,
\label{thetaflucdif}
\eeq
which diverges as $r$ (or, equivalently, time $t$), goes to infinity for $d<2$.
As often happens, the vanishing of this exponent $1-d/2$ in (\ref{thetaflucdif}) in $d=2$ indicates {\it not} a constant, but a logarithm: in fact, a slightly more careful version of the
reasoning used here, applied in exactly $d=2$, shows that
\beq
\sqrt{<\theta^2>}\propto\sqrt{\ln(r)}\to\infty \,.
\label{MW}
\eeq

So we've shown by this purely dynamical argument that, for $d\le2$, fluctuations diverge in the limit of an infinitely large system. Hence, there can be no long ranged order in our system of pointers for those spatial dimensions. This is the Mermin-Wagner theorem, derived in a very unorthodox dynamical way. In the final section of these notes, I'll show that modifying this argument to take into account motion shows that movers {\it can} order in $d=2$. But first, I'll show this more formally and systematically using hydrodynamics.

\newpage

    \section{Formulating the hydrodynamic model}

In this section,
I'll review the derivation and analysis of the
hydrodynamic model of polar ordered dry active fluids, which I'll also refer to as ``ferromagnetic  flocks". More details 
can be found in references\cite{TT3,TT4,NL}.  

As discussed in the introduction,
the system we wish to model is any collection of a large number $N$ of
organisms (hereafter referred to as ``birds'') in a
$d$-dimensional space, with each organism seeking to move in the
same direction as its immediate neighbors.

I further assume that each organism has no ``compass;'',   in
the sense defined in the Introduction, i.\ e.,
no intrinsically preferred direction in which it wishes to move.
Rather, it is equally happy to move in any direction picked by its
neighbors.  However, the navigation of each organism is not
perfect; it makes some errors in attempting to follow its
neighbors.  I consider the case in which these errors have zero
mean; e.\ g., in two dimensions, a given bird is no more likely to
err to the right than to the left of the direction picked by its
neighbors.  I also assume that these errors have no long
temporal correlations; e.\ g., a bird that has erred to the right at
time $t$ is equally likely to err either left or right at a time
$t^\prime$ much later than $t$.

The continuum model   will describe the
long distance behavior of {\it any} flock satisfying the
symmetry conditions I'll specify in a moment. The automaton
studied by Vicsek et al \cite{Vicsek} described in the introduction
provides one concrete realization of such a model.  Adding ``bells and
whistles'' to this model by, e.g.,  including purely attractive or repulsive
interactions between the birds, restricting their field of vision to those
birds ahead of them, giving them some short-term memory, etc., will
not change the hydrodynamic model, but can be incorporated simply
into a change of the numerical values of a few phenomenological
parameters in the model, in much the same way that
all simple fluids are described by the Navier-Stokes equations, and
changing fluids can be accounted for simply by changing, e.g.,  the viscosity
that appears in those equations.

This model should also describe real flocks of real living organisms,
provided that   the flocks are large enough, and that they have the same
symmetries and conservation laws that, e.g., Vicsek's algorithm does.

So, given this lengthy preamble, what {\it are} the
symmetries and conservation laws of flocks?

   The only symmetries of the model are invariance under rotations and
translations.
Translation-invariance simply means that displacing the positions of the
whole flock rigidly by a constant amount has no physical effect, since the
space the flock moves through is assumed to be on average homogeneous
\cite{transinv}.
Since I am not considering translational
ordering, this symmetry remains unbroken. Rotation invariance
simply says the ``birds'' lack a compass, so that all directions of space
are equivalent.  Thus, the ``hydrodynamic''
equation of motion
I write down cannot have built into it any special direction
picked ``a priori''; all directions must be spontaneously picked out
by the motion and spatial structure of the flock.  As we shall see,
this symmetry {\it severely} restricts the allowed terms
in the equation of motion.

Note that the model does {\it not} have Galilean
invariance:  changing the velocities of all the birds by some
constant boost $\bv_b$ does {\it not} leave the model
invariant.  Indeed, such a boost is {\it impossible} in a
model that strictly obeys Vicsek's rules, since the
{\it speeds} of all the birds will not remain equal to $v_0$
after the boost.  One could image relaxing this constraint on the
speed, and allowing birds to occasionally speed up or slow down,
while tending an average to move at speed $v_0$.  Then the
boost just described would be possible, but clearly would change
the subsequent evolution of the flock.

Another way to say this is that birds move through a resistive
medium, which provides a special Galilean reference frame, in
which the dynamics are particularly simple, and different from
those in other reference frames.  Since real organisms in flocks
always move through such a medium (birds through the air, fish
through the sea, wildebeest through the arid dust of the
Serengeti), this is a very realistic feature of the
model \cite{galinvfoot}.

As we shall see shortly, this {\it lack} of Galilean
invariance
{\it allows} terms in the hydrodynamic equations of birds
that are {\it not} present in, e.\ g., the Navier-Stokes
equations for a simple fluid, which {\it must} be Galilean
invariant, due to the absence of a luminiferous ether.

The sole conservation law for flocks is conservation of birds:  we
do not allow birds to be born or die ``on the wing''.

In contrast to the Navier-Stokes equation, I here consider systems without momentum, due to the presence of the resistive background medium which breaks  Galilean invariance.

Having established the symmetries and conservation laws
constraining our model, we need now to identify the
hydrodynamic variables.

What do I mean by ``hydrodynamic"?\cite{Forster} I mean variables that evolve slowly at long wavelength. More precisely, I mean variables whose evolution rate goes to zero as the length scale on which they are probed goes to infinity. 

When one first hears this concept, it is natural to wonder why there should be {\it any} such variables. For example, in a flock consisting of millions of organisms, wouldn't one expect all variables to relax on some ``microscopic" time scale, such as the mean time scale of interaction between neighboring birds?

This reasoning is {\it almost} correct: {\it almost} any variable one can think of in any system with an enormous number of degrees of freedom will relax back, on a microscopic time scale, to a value determined by the local values of the few ``slow", or ``hydrodynamic" variables. But again, why should {\it any} variable be slow?

 There are two possible reasons a variable will be slow: 
 
 \noindent 1) conservation laws, and
 
 \noindent 2) broken continuous symmetries. 
 
The density $\rho$ is an example of a variable which is hydrodynamic for the first reason. Variables that are slow for second reason  are called ``Goldstone modes". In our problem, rotation invariance implies that $\dv_\perp$, 
defined via
\beq
\bv(\br,t)=v_{0}\hat{x}_{\parallel}+\dv(\br,t)=(v_{0}+\dvpa)\hat{x}_{\parallel}+\dv_\perp(\br,t)
\label{vexp}
\eeq
is a hydrodynamic variable. This is because a constant $\dv_\perp$ amounts to just a rotation, if $\dvpa$ relaxes back to the value $\dvpa=\sqrt{v_0^2-|\dvp|^2}-v_0$ required to keep $|\bv|=v_0$. Since the system is rotation invariant, such a spatially uniform variation of $\dv_\perp$ can {\it never} relax; i.e., it has an infinite lifetime. Therefore, by continuity, if the field $\dv_\perp$ varies slowly in space, it must relax very slowly. More precisely, the relaxation time of such a distortion in $\dv_\perp$ must go to infinity as the length scale on which it varies does.

We've already seen an illustration of this for the pointer problem: as distance $r\to\infty$, the time $t$ required for the field $\theta$ to equilibrate over that distance $r$ diverges like $r^2$. This is because $\theta$ is the Goldstone mode, in the sense just described, for the pointer problem. The broken continuous symmetry with which $\theta$ is associated- that is, the symmetry that guarantees that $\theta$ will be ``slow" at long wavelengths (which is precisely what I mean by   ``hydrodynamic") is just rotation invariance.

Note that although $\dvp$ is a hydrodynamic variable, $\dvpa$ is not, since there is no symmetry that forbids the {\it speed} of the flockers from relaxing back to the preferred speed $v_0$ in a finite time, {\it even if} the fluctuation of the speed  away from $v_0$ is spatially uniform. Nonetheless, because it is far simpler to see the consequences of rotation nvariance for the full velocity field $\bv$ than it is for the perpendicular component $\dvp$ of $\bv$ alone, I will initially formulate  hydrodynamic equations of motion for the full velocity $\bv$, even though this will include the non-hydrodynamic variable $\dvpa$. Once I have the equations of motion, it is then conceptuallly straightforward (although algebraically fairly monstrous, as we'll see) to eliminate $\dvpa$ and rewrite the equations of motion entirely in terms of the hydrodynamic variables  $\dvp$ and $\rho$.

I will also follow the historical precedent of the
Navier-Stokes\cite{Forster},\cite{FNS} equation by deriving our
continuum, long wavelength description of the flock {\it not} by explicitly
coarse graining the microscopic dynamics (a
{\it very} difficult procedure in practice), but, rather, by
writing down the most general continuum equations of motion for
$\bv$ and $\rho$ consistent with the symmetries and
conservation laws of the problem.  This approach allows us to
bury our ignorance in a few phenomenological parameters, (e.\ g.,
the viscosity in the Navier-Stokes equation) whose numerical
values will depend on the detailed microscopic rules of individual
bird motion.  What terms can be present in the EOMs, however,
should depend only on symmetries and conservation laws, and
{\it not} on   other aspects of the microscopic rules.

To reduce the complexity of our equations of motion still further,
I will perform a spatio-temporal gradient expansion, and keep
only the lowest order terms in gradients and time derivatives of
$\bv$ and $\rho$.  This is motivated and justified by our
desire to consider {\it only} the long distance, long time
properties of the flock.  Higher order terms in the gradient
expansion are ``irrelevant'':  they can lead to {\it finite}
``renormalization'' of the phenomenological parameters of the
long wavelength theory, but {\it cannot} change the type
or scaling of the allowed terms.

So let's begin.

Rotation invariance implies that $\pp_t\bv$, being a vector itself, must equal a sum of some other vectors. So, what vectors can we make out of $\bv$, the scalar $\rho$, and the gradient operator?

\begin{figure}
 \includegraphics[width=1.0\textwidth]{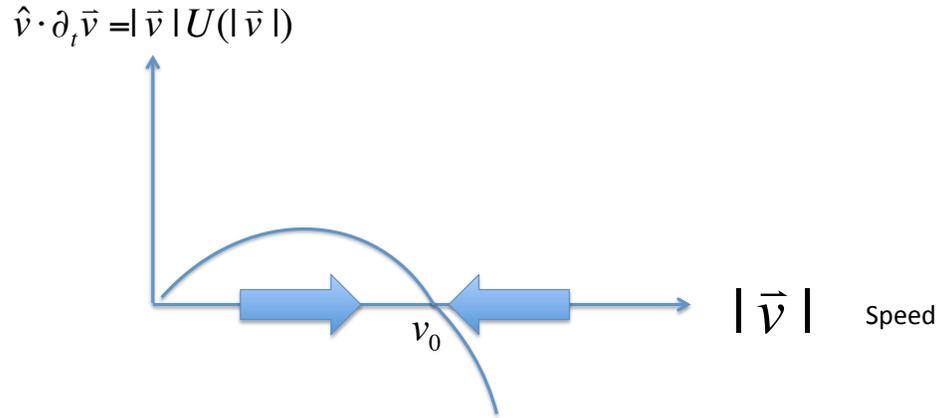}
 \caption{\label{U}Plot of the acceleration along $\bv$ arising from the $\bv U(|\bv|)$ term in the equation of motion. This is the simplest qualitative form that can lead (for sufficiently small noise) to an ordered, moving flock.}
\end{figure}

Well, the most obvious vector is $\bv$ itself. More generally, we can multiply $\bv$ by any scalar function of the speed $|\bv|$ and the density  $\rho$:

\beq
(\pp_t\bv)_1=U(|\bv|, \rho)\bv
\label{Uterm}
\eeq
This looks like a conventional frictional drag coefficient, except for the crucial difference that, while the drag coefficient must always be negative (friction slows down a passive particle), for an active system, we'll allow $U>0$, at least for small $|\bv|$.  This is how we make our system active. We don't want $U$ to be positive for all $|\bv|$; if it was, the speed of the flock would grow without bound, which is clearly unphysical. So we will assume that $U$, plotted as a function of the speed $|\bv|$, is positive for small speeds $|\bv|$, and turns negative for large speeds $|\bv|$. This leads to the acceleration in the direction of  motion  illustrated in figure (\ref{U}).

The effect of such a form for $U$ is clearly the following:  an initially slowly moving flock (or region thereof) will increase its speed until it reaches the speed $v_0$ at which $U(|\bv|)$ vanishes. Likewise, a flock (or region thereof) that is moving faster than $v_0$ will slow down until its speed again reaches $v_0$. Thus the speed $|\bv|$ is {\it not} a hydrodynamic variable; it relaxes back in a finite time $\tau=1/(\pp U/\pp|\bv|)_\rho$ to $v_0$.

There are, obviously, infinitely many functions of the speed $|\bv|$ and $\rho$ that have the properties just described. Fortunately, since the speed always adjusts itself to be close to $v_0$, there prove to be only three parameters that we need to extract from $U$ for our hydrodynamic theory:
the steady state speed  $v_0$,  and the derivatives $\pp U/\pp|\bv|$ and  $\pp U/\pp\rho$ evaluated at $|\bv|=v_0$ and $\rho=\rho_0$, where $\rho_0$ is the mean density.

\begin{figure}
 \includegraphics[width=1.0\textwidth]{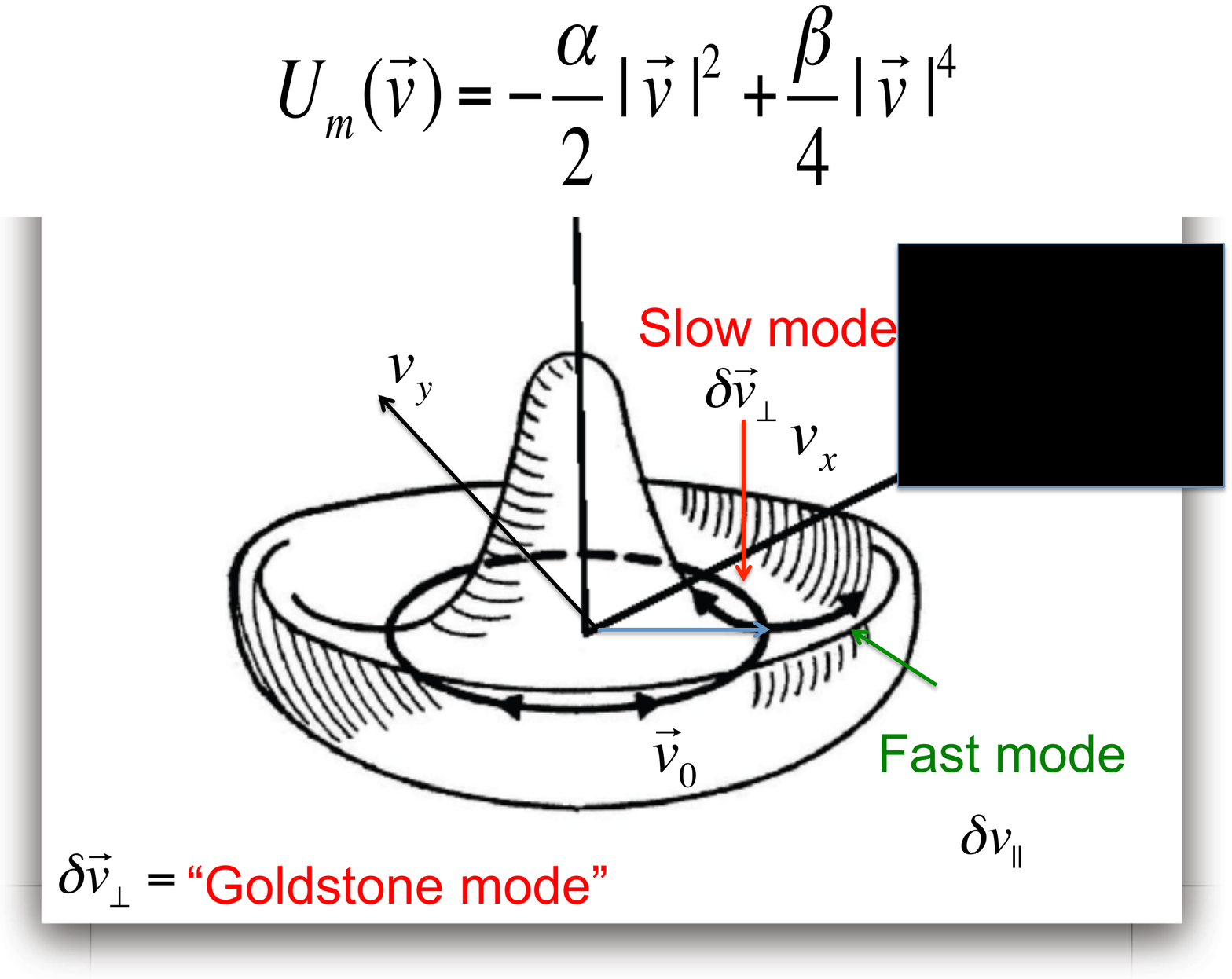}
 \caption{\label{V}The ``Mexican hat" potential. Note the circular ring of minima. Fluctuations $\dvp$ of the velocity that  move it aroiund this ring are ``Goldstone modes", which experience no restoring force from this ``potential". In contrast, fluctuations $\dvpa$ that carry one up the brim of the hat, or towards the crown (i.e., radially), are ``fast"; i.e., they will relax quickly due to this potential, which simply reflects the existence of a preferred speed for the flockers' motion through the frictional medium.}
\end{figure}

One popular choice for the function $U$ (indeed, the choice Yuhai and I made in our early papers on this problem) is the ``$\psi^4$" theory:
\beq
U=\alpha(\rho)-\beta(\rho)
|\bv|^{2} \,.
\label{psi4}
\eeq
The reason this is called 
``$\psi^4$" theory is that with this choice, we can write
\beq
U\bv=-\pp V(\bv)/\pp\bv \,,
\label{VU}
\eeq
where the ``potential" 
\beq
V(\bv)=-{1\over2}\alpha(\rho)|\bv|^{2}+{1\over4}\beta(\rho)
|\bv|^{4}
\label{Vdef}
\eeq
takes the form of the famous ``Mexican hat", as shown in figure (\ref{V}).  The dynamical effect of the $U(|\bv|)$ term is then simply to make the velocity $\bv$ evolve down towards the circular (or, in three dimensions, spherical) ring of identical minima at $|\bv|=v_0=\sqrt{\alpha/\beta})$.

This form is widely used in Condensed matter physics and field theory. It is also the form most appropriate for studying the order-disorder transition. However, since here I'm just interested in the behavior of the flock deep inside its ordered phase, I will not restrict myself to this form. It is useful, however, to keep figure (\ref{V}) in mind, as we can construct a potential via (\ref{VU}) for any $U$, and, if that $|\bv|U$ has the form plotted in figure (\ref{U}), the associated $V$ will look qualitatively like figure (\ref{V}). And we can imagine the flock velocity $\bv$  evolving by seeking the minimum of this potential, or, more precisely, the ring of minima of this potential. 

Indeed, the ``spontaneous symmetry breaking" of a flock can be thought of as the system's settling into one of these degenerate minima.

Note that if we use the expansion  (\ref{vexp}) around the mean velocity in  (\ref{Uterm}), 
 we get no linear term in $\dv$ in the equation of motion for $\dvp$. To see this, note that the speed
 \beq
 |\bv|=\sqrt{(v_{0}+\dvpa)^2+|\dv_\perp(\br,t)|^2}=v_{0}+\dvpa+\cO(|\dv_\perp(\br,t)|^2)\,.
 \label{speed exp}
 \eeq
 Using this in the projection of (\ref{Uterm}) perpendicular to the mean velocity, we obtain
 \begin{eqnarray}
(\pp_t\dvp)_1&=&U(v_0+\dvpa, \rho_0)\dvp+\cO(|\dvp|^3)\nonumber\\&=&U(v_0, \rho_0)\dvp+\left(\pp U/\pp|\bv|\right)|_{v_0, \rho_0}\dvpa\dvp+\left(\pp U/\pp\rho\right)|_{v_0, \rho_0}\delta\rho\dvp+\cO(|\dvp|^3)\nonumber\\
&=&U(v_0, \rho_0)\dvp+\cO(|\dvp|^3, \dvpa\dvp,\delta\rho\dvp)=\cO(|\dvp|^3, \dvpa\dvp,\delta\rho\dvp)\,,
\label{Uterm exp}
\end{eqnarray}
where in the last equality I have used the fact that $U(v_0, \rho_0)=0$, which is just a consequence of the definition of $v_0$ as the steady state speed.

This vanishing of all terms linear in $\dv$ in this term is no coincidence; rather, it is a consequence of the fact that $\dvp$ is the Goldstone mode for this problem, so any term in it's time derivative {\it must} have some spatial gradients, so that it vanishes when $\dvp$ is spatially uniform.

  This means if we want to include terms that depend on $\dv_\perp$ (which we certainly do!), then we need to look at terms involving the gradient operator ${\bf\nabla}$.
  
So let's look at:

\noindent 2) Combinations of velocities and one gradient operator. We need at least two velocities. Why? Well,  can we make anything with a single gradient operator and a single velocity that transforms like a vector? The answer is no, as can most easily be seen using the Einstein summation convention: if we write
\beq
(\pp_tv_i)_{\rm trial}={\rm constant} \times\pp_j v_k
\label{1v1grad}
\eeq
there's no choice of the indices j and k on the right hand side that will make this equation make sense. If we take $j=i$, we have an extra index $k$ running loose on the right hand side. If we try to get rid of it by instead taking $j=k$, then we've made the right hand side into a scalar (in fact, into $\bna\cdot\bv$), which we can't equate to a vector. So we need at least two velocities to combine with our gradient operator.

So let's try
\beq
(\pp_tv_i)_{\rm 2}={\rm constant} \times v_\ell\pp_j v_k
\label{2v1grad}
\eeq
This can be made to work with a suitable choice of the three indices $\ell$, $j$, and $k$. What we need to do is make two of  them equal, so that they are summed over by the Einstein summation convention. This leaves one free index, which we must choose to be $i$, the free index on the right hand side. Basically, we use the Einstein summation convention to ``eat" the extra indices on the right hand side. In fact, there are three ways to make this work:

\noindent i) Take $k=i$ and $\ell=j$. The right hand side has only one free index, so it's a vector, and it's the same free index as on the left hand side, so the equation makes sense in the Einstein summation convention. We can write this term as

\beq
(\pp_tv_i)_{\rm 2.1}=-\lambda_1  \bv \cdot\bna v_i \,,
\label{lambda1def2}
\eeq
where I've arbitrarily defined the ``constant" in (\ref{2v1grad}) to be $-\lambda_1$. (The minus sign is chosen to make the resulting equation look as much like the Navier-Stokes equation as possible, as you'll see).
Rewriting this in full glorious vector notation,
\beq
(\pp_t\bv)_{\rm 2.1}=-\lambda_1  \bv \cdot\bna \bv \,.
\label{lambda1def2}
\eeq

\noindent ii) Take $k=j$ and $\ell=i$. Once again, the right hand side has only one free index, so it's a vector, and its the same free index as on the left hand side, so the equation makes sense in the Einstein summation convention. We can write this term as

\beq
(\pp_tv_i)_{\rm 2.2}=-\lambda_2  v_i \bna\cdot\bv \,,
\label{lambda2def}
\eeq
where I've called the ````constant" $-\lambda_2$.  In vector notation,
\beq
(\pp_t\bv)_{\rm 2.2}=-\lambda_ 2 \bv (\bna \cdot \bv) \,.
\label{lambda2def}
\eeq

\noindent iii) take $j=i$ and $\ell=k$. This also makes sense in the Einstein summation convention. We can write this term as

\beq
(\pp_tv_i)_{\rm 2.3}=-2\lambda_3   v_j\pp_i v_j =-\lambda_3   \pp_i (|\bv|^2)\,,
\label{lambda3def}
\eeq
where I've introduced the factor of $2$ in this definition of $\lambda_3$ for convenience in writing the second equality. In vector notation,
\beq
(\pp_t\bv)_{\rm 2.3}=-\lambda_3  \bna (|\bv|^2) \,.
\label{lambda3def2}
\eeq

There are also combinations of one gradient, $\bv$, and the density $\rho$ that do not have the structure of (\ref{2v1grad}), particularly if we allow more than one power of $\bv$. Note that there is no reason we should not include such higher powers: because of the spontaneous ordering, $|\bv|$ itself is not small (in fact, it's close to $v_0^2$, which need not be small). We {\it do} intend to expand in powers of the fluctuation $\dv$ of $\bv$ away from its mean value $v_0\hx$, but that is {\it not} the same as an expansion in powers of $\bv$, because of this spontaneous order.

Fortunately, it turns out that we can incorporate all such one gradient terms into five terms, namely:

\noindent I) a pressure term 
\beq
(\pp_t\bv)_{\rm pressure}=-  \bna P(|\bv|, \rho) \,.
\label{lpressuredef}
\eeq
Those of you familiar with the Navier-Stokes equation will recognize this as exactly the form of the pressure term in that equation, except for the peculiarity here that the pressure can depend not only on the density $\rho$, but also on the speed $|\bv|$. Such dependence is forbidden in the Navier-Stokes equation by Galilean invariance; since we don't have Galilean invariance in our dry active fluid, this dependence is allowed, and hence will, in general, be present.

II-IV) density and speed dependences of the $\lambda_{1,2,3}$ terms, and

V) an anisotropic pressure term $P_2$ of the form 
\beq
(\pp_t\bv)_{\rm aniso \, pressure}=-\bv(\bv\cdot\bna)P_2(|\bv|, \rho)
\label{P2def}
\eeq

To see that these five terms exhaust all possibilites, consider, for example, the term
\beq
(\pp_tv_i)_{3}=v_\ell v_k\pp_j f(|\bv|, \rho)
\label{towardsP2}
\eeq
Again choosing the indices so that two of them are eaten by the Einstein summation convention, while the remaining one is $i$, we see that there are two ways to do this:

\noindent i) $j=i$, $k=\ell$. This choice gives 
\beq
(\pp_tv_i)_{3.1}=|\bv|^2 \bna f=\bna (|\bv|^2 f)-f \bna v^2\equiv -\bna\delta P-\delta\lambda_2\bna v^2 \,.
\label{Pressure}
\eeq
where I've defined a contribution $\delta P(|\bv|, \rho)\equiv f(|\bv|, \rho)|\bv|^2$ to the ``Pressure" defined above, and a contribution $\delta\lambda_2(|\bv|, \rho)\equiv f(|\bv|, \rho)$ to $\lambda_2(|\bv|, \rho)$.

\noindent ii) $j=k$, $\ell=i$ (note that $j=\ell$, $k=i$ gives the same term). 
\beq
(\pp_tv_i)_{3.2}=v_iv_j\pp_j f(|\bv|, \rho) \,,
\label{P2def}
\eeq
which in vector form is precisely the  $P_2$ term (\ref{P2def}) with $P_2(|\bv|, \rho)=-f(|\bv|, \rho)$.

The $\lambda_{1,2,3}$, $P$,  and $P_2$ terms can between them incorporate every ``relevant" (i.e., non-negligible)  term that involves one gradient and arbitrary powers on $\bv$ and $\rho$. To see this, consider, 
for example,  the following term with four velocities and one gradient:
\beq
(\pp_tv_i)_{{\rm trial}2}=\lambda_4  v_\ell\pp_j (v_nv_mv_k)
\label{4v1grad}
\eeq
We need to ``eat"  four of the five indices on the right hand side, and set the remaining one equal to $i$. Let's consider the term we get if we choose $m=n$, $\ell=j$, and $k=i$. This gives
\beq
(\pp_tv_i)_{4}=\lambda_4 \bv\cdot\bna (v_i|\bv|^2)=\lambda_4[|\bv|^2\bv\cdot\bna (v_i)
+v_i \bv\cdot\bna |\bv|^2] \,.
\label{4v1grad2}
\eeq

The first term on the right hand side is immediately recognizable as a contribution to $\lambda_1$ proportional to $|\bv|^2$, while the second is a constant contribution to $P_2$. 

It is straightforward to check that all terms that involve only one gradient can likewize be incorporated into speed $|\bv|$ and density $\rho$- dependent corrections to one of the five aforementioned quantities isotropic pressure $P$, anisotropic pressure $P_2$, and $\lambda_{1,2,3}$.

\noindent 3) Let's now consider terms with two gradients. One might think that we need not keep such terms, since they have more gradients than the  one gradient terms we've just considered. However, it turns out, as we'll see in the next section, that none of the one gradient terms we've just considered damps out velocity fluctuations to linear order in the velocity fluctuations $\dv$, which prove to be small. Instead, they just lead to propagation without dissipation. Therefore, if we do not include any two gradient terms, our theory  would (erroneously) predict that there would be no damping of the fluctuations induced by the noise, which would therefore grow without bound over time. To prevent such an unphysical result, we need to go to higher order gradient terms. Second order proves, again with hindsight, to be sufficient.

So what can we make with two gradients that transforms like a vector? As before, let's proceed by writing out possible terms in Einstein summation convention, and figure out how the indices can get eaten. So let's start with terms with one velocity and two gradients. Generically, this can be written:
\beq
(\pp_tv_i)_{\rm trial 3}={\rm constant} \times\pp_j\pp_k v_\ell
\label{1v2grad}
\eeq

By now, you should be familiar enough with how this goes to see that there are two menu options for "index eating":

\noindent i) $j=k$ and $\ell=i$. This gives
\beq
(\pp_tv_i)_{4.1}=D_{T} \nabla^2 v_i
\label{shear tensor}
\eeq
or, in vector notation,
\beq
(\pp_t\bv)_{4.1}=D_{T} \nabla^2 \bv
\label{shear vector}
\eeq

\noindent ii) $j=i$ and $\ell=k$ (or, equivalently, $k=i$ and $\ell=j$), which gives

\beq
(\pp_tv_i)_{4.2}=D_{B} \pp_i\pp_j v_j
\label{bulk tensor}
\eeq
or, in vector notation,
\beq
(\pp_t\bv)_{4.2}=D_{B} \bna(\bna\cdot \bv)
\label{bulk vector}
\eeq

That's it for terms with two spatial gradients and one velocity. These two terms also occur in the Navier-Styokes equations, where the coefficients $D_{T}$ and $D_{B}$ are usually denoted as $\nu_{T}$ and  $\nu_{B}$, and are called the shear and bulk viscosities, respectively.

Can we make terms with more velocities and two derivatives? Absolutely; indeed, an overabundance of them. We can, however, tremendously reduce the number of possibilities by noting (as we did for the one gradient terms above)  that when we expand about the state of uniform motion via (\ref{vexp}), any velocity that a gradient acts on can be replaced by $\dv$, since $v_0\hxp$ is a constant. Since $\dv$ is small, the dominant terms will be those with only one $\dv$. We can therefore restrict ourselves to terms with only one full velocity $\bv$ acted upon by the two derivatives. Therefore, all possible ``relevant" terms involving two gradients and an {\it arbitrary} number of velocities can be written

\beq
(\pp_tv_i)_{4\,{\rm general}}={\rm constant} \times[v_pv_n\cdot\cdot\cdot v_sv_u]\pp_j\pp_k v_\ell
\label{manyvs1grad}
\eeq
where $[v_pv_n\cdot\cdot\cdot v_sv_u]$ is a product of  an even number $2m$ of  components of $\bv$. This number must be even,   so that there are an odd number of indices altogether on the right hand side. This is necessary to allow us to pair all but one of them off, thereby producing a vector.
 There are now four ways we can do this pairing off:
 
\noindent i) Pair all of the $v$'s to the left of the derivatives off with themselves, and set $j=k$ and $\ell=i$. This gives
\beq
(\pp_tv_i)_{4.1}={\rm constant} \times |\bv|^{2m}\pp_j\pp_j v_i={\rm constant} \times |\bv|^{2m}\nabla^2 v_i
\label{DT cont tensor}
\eeq
or, in vector notation,
\beq
(\pp_t\bv)_{4.1}={\rm constant} \times |\bv|^{2m}\nabla^2 \bv
\label{DT cont vector}
\eeq
We can absorb this into a contribution to the ``shear viscosity" $D_{T}$ proportional to $|\bv|^{2m}$. We can therefore incorporate all possible such terms, up to arbitrary even powers of $|\bv|$, by making $D_{T}$ a suitably chosen function of $|\bv|$. We can generalize this even further by making $D_{T}$ depend on the density $\rho$ as well.

\noindent ii) Pair all of the $v$'s to the left of the derivatives off with themselves, and set  $j=i$ and $\ell=k$ (or, equivalently, $k=i$ and $\ell=j$). This gives
\beq
(\pp_tv_i)_{4.2}={\rm constant} \times |\bv|^{2m}\pp_i\pp_j v_j
\label{DB cont tensor}
\eeq
or, in vector notation,
\beq
(\pp_t\bv)_{4.2}={\rm constant} \times |\bv|^{2m}\bna(\bna\cdot \bv)
\label{DB cont vector}
\eeq
which we can absorb into a contribution to  the ``bulk visocosity" $D_{B}$ proportional to $|\bv|^{2m}$. We can therefore incorporate all possible such terms, up to arbitrary even powers of $|\bv|$, by making $D_{B}$ a suitably chosen function of $|\bv|$. As for $D_{T}$, we can generalize this even further by making $D_{B}$ depend on the density $\rho$ as well.

\noindent iii) Pair all but two of the $v$'s to the left of the gradients off with themselves, and pair the remaining two t with  the gradients; this forces $\ell=i$ (since there are no other free indices left).   This gives
\beq
(\pp_tv_i)_{4.3}={\rm constant} \times |\bv|^{2m-2}v_jv_k\pp_j\pp_k v_i
\label{D2 cont tensor}
\eeq

or, in vector notation,
\beq
(\pp_t\bv)_{4.3}={\rm constant} \times |\bv|^{2m-2}(\bv\cdot\bna)^2 \bv
\label{DB cont vector}
\eeq
This is the first genuinely new term. I'll sum up all such terms into a function that I'll call $D_2(|\bv|, \rho)$ of the speed $|\bv|$ and the density $\rho$ times the combination $(\bv\cdot\bna)^2 \bv$. This term makes anisotropic diffusion possible: we can now have a different diffusion constant along the direction of flock motion than perpendicular to it, as we would expect, since we've broken (or, rather, the flock has broken) the symmetry 
between the direction of flock motion and directions perpendicular to it.

\noindent iv) Pair all but two of the $v$'s to the left of the gradients off with themselves, and pair the one of the other with one of the gradients, and the other with the velocity to the {\it right} of the gradients; this forces one of the gradient indices to be $i$ (since there are no other free indices left).  This gives
\beq
(\pp_tv_i)_{4.4}={\rm constant} \times |\bv|^{2m-2}v_jv_k\pp_i\pp_k v_j
\label{1st irrel cont tensor}
\eeq
This contribution proves to be negligible compared to those we've already kept. To see this, consider the implied sum on $j$ in (\ref{1st irrel cont tensor}). One term in this sum is that with index $j=\parallel$; i.e., the Cartesian component along the mean direction of motion. We can replace $v_\parallel$ with $\delta v_\parallel$ to the right of the gradient in (\ref{1st irrel cont tensor}), since the mean velocity contribution to this term $v_0\hx$ is a onstant, and hence has zero gradient. But, as we noted earlier, $\delta v_\parallel\ll|\bvp|$ since $\delta v_\parallel$ is not a Goldstone mode, so this contribution from the sum on $j$ to this term is negligible compared to the two gradient, one $\bvp$ terms we found above.

The other terms in the sum on $j$ will be proportional to two $\bvp$'s (one to the left of the gradient, and one to the right), and so will be negligible compared to the ``one $\bvp$, two gradient" terms found above if $\bvp$ is small, as it will be in an ordered state. So those terms in the sum are negligible as well. So the entire term (\ref{1st irrel cont tensor}) is negligible.

\noindent v) Finally, we can pair all but {\it four} of the $v$'s to the left of the gradients among themselves, set one of the remaining indices on those $v$'s equal to $i$, and pair off the remaining three velocities with the two gradients and the velocity to the right of the gradient. This gives
\beq
(\pp_tv_i)_{4.5}={\rm constant} \times |\bv|^{2m-4}v_iv_jv_kv_l\pp_j\pp_k v_\ell
\label{2nd irrel cont}
\eeq
The sum on $\ell$ in this term can be shown to be negligible by an argument almost identical to the one we just used for the previous contribution (\ref{1st irrel cont tensor}). So we'll drop this as well.

The only other term we need to include is a random noise term:
\beq
(\pp_t\bv)_{noise}=\bff(\br,t)
\label{noise}
\eeq
 It is  assumed to be Gaussian with
white noise correlations:
\beq
\langle f_{i}(\bbr,t)f_{j}(\bbr',t')\rangle=2D
\delta_{ij}\delta^{d}(\bbr-\bbr')\delta(t-t')
\label{white noise}
\eeq
where the ``noise strength" $D$ is a constant parameter of the system, and $i,j$ denote
Cartesian components.

Using the dynamical RG, one can show that small departures of the noise statistics from
 purely Gaussian have no effect on the long-distance physics.

Dah-deeb, dah-deeb, that's all, folks! Any other terms you construct will have  more gradients, and so will be negligible at long distances compared to the terms we've already found.

Putting all of these terms together gives the equation of motion for $\bv$:
\begin{eqnarray}
&&\partial_{t}
\bv+\lambda_1(\bv\cdot{\bf\nabla})\bv+
\lambda_2({\bf\nabla}\cdot\bv)\bv
+\lambda_3{\bf\nabla}(|\bv|^2) = \nonumber \\
&&U(|\bv|, \rho)\bv -{\bf\nabla} P +D_{B} {\bf\nabla}
({\bf\nabla}
\cdot \bv)
+ D_{T}\nabla^{2}\bv +
D_{2}(\bv\cdot{\bf\nabla})^{2}\bv+\bff
\label{EOM}
\end{eqnarray}

Keep in mind that this equation is (even!) more complicated than it looks, because all of the parameters $\lambda_i (i = 1 \to 3)$,
$U$, the ``damping coefficients" $\mu_{B,T,2}$, the  ``isotropic pressure'' $P(\rho,
v)$ and the  ``anisotropic Pressure'' $P_2 (\rho, v)$
are  functions of the density $\rho$ and the
magnitude $v\equiv|\bv|$ of the local velocity. 

To close these equations of motion, we also need one for the density. The final
equation (\ref{conservation}) is just conservation of bird number (we
don't allow our birds to reproduce or   die on the wing).

\begin{eqnarray}
{\partial\rho \over \partial
t}+{\bf\nabla}\cdot(\bv\rho)=0
\label{conservation}
\end{eqnarray}

With these equations of motion (\ref{EOM}) and (\ref{conservation}) in hand, we can now use them to figure out how flocks actually behave, and  in particular why  they can order even in $d=2$.

\section{Solving the hydrodynamic model}
\subsection{Expanding the equations of motion to ``relevant" non-linear order}
The hydrodynamic model embodied in equations (\ref{EOM}) and (\ref{conservation})  is equally valid  in both the
``disordered '' (i.e., non-moving) state, in which $U(|\bv|, \rho)$ is negative for all $|\bv|$, and in the moving or 
``ferromagnetically ordered' state, in which $|\bv|U(|\bv|)$ looks like figure (\ref{U}), with a positive region at small $|\bv|$, which allows for the possibility of a moving state. In
this section I'll focus on the ``ferromagnetically
ordered'', broken-symmetry phase; and specifically on the question of whether fluctuations
around the symmetry broken ground state destroy the ordered phase (as in the analogous
phase of the 2D XY model). When $|\bv|U(|\bv|)$ looks like figure (\ref{U}), we can write the expand the velocity field as in (\ref{vexp}), which I rewrite here for convenience:
\beq
\bv(\br,t)=v_{0}\hat{x}_{\parallel}+\dv(\br,t)=(v_{0}+\dvpa)\hat{x}_{\parallel}+\dv_\perp(\br,t)\,,
\label{vexp2}
\eeq
where I remind you that 
$v_{0}\hat{x}_{\parallel}$ is the spontaneous average
value of
$\bv$ in the ordered phase in the absence of fluctuations, whose magnitude $v_0$  is just that at which $U(|\bv|, \rho)=0$.

As I've discussed above, the fluctuation $\dvpa$ of the component of $\bv$ along the mean direction of flock motion $\hat{x}_{\parallel}$ away from its preferred value $v_0$ is {\it not} a hydrodynamic variable of the system; rather, it relaxes back quickly to a value determined by the true hydrodynamic variables $\rho$ and $\dvp$. It therefore behooves us to eliminate it by {{\cred solving for it in terms of} those variables. Doing so is rather tricky - indeed, Yuhai and I got this slightly wrong in our earlier work on this problem \cite{TT1,TT2,TT3,TT4}- so I will go through the argument rather carefully and in some detail here. For further details, see\cite{NL}.

Since we know fluctuations in the speed (i.e., the magnitude $|\bv|$ of $\bv$) will be fast, it is useful to turn our equation of motion (\ref{EOM}) for the velocity into an equation of motion for that speed. 
This can be done by 
taking the dot product of both sides of equation (\ref{EOM}) with $\bv$ itself, which gives:
\begin{widetext}
\begin{eqnarray}
{1\over 2}\left(\partial_{t}|\bv|^2+(\lambda_1 + 2 \lambda_3)(\bv\cdot{\bf\nabla})|\bv|^2\right) &+& \lambda_2({\bf\nabla}\cdot\bv)|\bv|^2= U(|\bv|,\rho)|\bv|^{2}-\bv \cdot {\bf\nabla}  P-|\bv|^{2}\bv \cdot {\bf\nabla}  P_2 \nonumber \\&+&D_{1} \bv\cdot{\bf\nabla}
({\bf\nabla}\cdot \bv) + D_{T}\bv\cdot\nabla^{2}\bv +D_{2}\bv\cdot\left((\bv\cdot{\bf\nabla})^{2}\bv\right)+\bv\cdot\bff\nonumber\\
\label{v parallel elim}~.
\end{eqnarray}
\end{widetext}
In this hydrodynamic approach, 
we are interested only in fluctuations $\dv(\br, t)$ and $\delta \rho(\br, t)$
that vary slowly in space and time. (Indeed, the hydrodynamic equations (\ref{EOM}) and (\ref{conservation}) are only valid in this limit). Hence, terms involving space and time derivatives of 
$\dv(\br, t)$ and $\delta \rho(\br, t)$
are always negligible, in the hydrodynamic limit, compared to terms involving the same number of powers of fields without any time or space derivatives.

Furthermore, the fluctuations 
$\dv(\br, t)$ and $\delta \rho(\br, t)$ can themselves be shown to be small in the long-wavelength limit. Hence, we need only keep terms in equation (\ref{v parallel elim}) up to linear order in 
$\dv(\br, t)$ and $\delta \rho(\br, t)$. The 
$\bv\cdot\bff$ term can likewise be dropped, since it only leads to a term of order 
$\bv_{_\perp} f_{_\parallel}$ in the $\bv_{_\perp}$ equation of motion, which is negligible (since $\bv_{_\perp}$ is small) relative to the $\bff_\perp$ term already there.

These observations can be used to eliminate many of the terms in equation (\ref{v parallel elim}), and solve for $
U$; 
the  solution  is:
\begin{eqnarray}
U=\lambda_2 {\bf\nabla}\cdot\bv+\bv\cdot{\bf\nabla}P_2+{\sigma_1\over v_0}\partial_{_\parallel} \delta \rho+{1\over 2 v_0}\left(\partial_t + \gamma_2 \partial_{_\parallel} \right)\delta v_{_\parallel} , \nonumber\\
\label{Usol}
\end{eqnarray}
where I've defined 
\begin{eqnarray}
\gamma_2\equiv(\lambda_1+2\lambda_3)v_0~
\label{gamma_2def}
\end{eqnarray}
and
\beq
\sigma_1\equiv\left({\pp P\over\pp\rho}\right)_0 \,.
\label{sigma1def}
\eeq
Here and hereafter , super-
or sub-scripts
$0$ denoting functions of  $\rho$ and 
$|\bv|$ evaluated at the steady state values $\rho = \rho_0$ and $ |\bv|=v_0$.

Inserting this expression (\ref{Usol}) for $U$ back into equation (\ref{EOM}), I find that $P_2$ and 
$\lambda_2$ cancel out of the 
$\bv$ equation of motion, leaving
\begin{widetext}
\begin{eqnarray}
\partial_{t}
\bv+\lambda_1(\bv\cdot{\bf\nabla})\bv+\lambda_3 {\bf\nabla}(|\bv|^2)
&=&{\sigma_1\over v_0}
\bv (\partial_{_\parallel} \delta \rho)-{\bf\nabla} P + D_{1} {\bf\nabla}
({\bf\nabla}
\cdot \bv)+ D_{T}\nabla^{2}\bv +
D_{2}(\bv\cdot{\bf\nabla})^{2}\bv\nonumber\\&+&\left[{1\over 2 v_0}\left(\partial_t + \gamma_2 \partial_{_\parallel} \right)\delta v_{_\parallel}
\right]\bv+\bff~~.
\label{EOM2}
\end{eqnarray}
\end{widetext}
This can be made into an equation of motion for $\bv_{_\perp}$ involving only $\bv_{_\perp}(\br, t)$ and $\delta \rho(\br, t)$ by projecting perpendicular to the direction of mean flock motion $\hat{x}_{_\parallel}$, and eliminating $\delta v_{_\parallel}$ using equation (\ref{Usol}) and 
the expansion 
\begin{eqnarray}
U\approx-\Gamma_1\left(\delta v_{_\parallel} +{|\bv_{_\perp}|^2\over 2 v_0}\right) - \Gamma_2 \delta \rho ~~,
\label{Uexp}
\end{eqnarray}
where 
I've defined 
\begin{eqnarray}
\Gamma_1 \equiv -\left({\partial U
 \over \partial |\bv|}\right)^0_{\rho} &,
&\Gamma_2 \equiv - \left({\partial U
 \over \partial \rho}\right)^0_{|\bv|}~.
\label{gamma12 def}
\end{eqnarray}
I've also used the expansion
(\ref{vexp2}) for the velocity in terms of the fluctuations $\delta v_{_\parallel}$ and $\dv_\perp$ to write
\begin{eqnarray}
|\bv|=v_0+\delta v_{_\parallel} +{|\bv_{_\perp}|^2\over 2 v_0}+O(\delta v_{_\parallel} ^2, |\bv_{_\perp}|^4)~,
\label{speed}
\end{eqnarray}
and kept only terms that an RG analysis shows to be relevant in the long wavelength limit.
Inserting  (\ref{Uexp}) into (\ref{Usol})  gives:

\begin{widetext}
\begin{eqnarray}
-\Gamma_1\left(\delta v_{_\parallel} +{|\bv_{_\perp}|^2\over 2 v_0}\right) - \Gamma_2 \delta \rho =\lambda_2 {\bf\nabla}_{_\perp}\cdot\bv_{_\perp}+\lambda_2\partial_{_\parallel} \delta v_{_\parallel}+{(\mu_1 v_0^2+\sigma_1)\over v_0}\partial_{_\parallel} \delta \rho+{1\over 2 v_0}\left(\partial_t + \gamma_2 \partial_{_\parallel} \right)\delta v_{_\parallel}~~,\nonumber\\
\label{v par 1}
\end{eqnarray}
\end{widetext}
where I've kept only linear terms on the right hand side of this equation, since the non-linear terms are at least of order derivatives of $|\bv_{_\perp}|^2$, and hence negligible, in the hydrodynamic limit, relative to the  $|\bv_{_\perp}|^2$ term explicitly displayed on the left-hand side.

This equation can be solved iteratively for $\delta v_{_\parallel}$ in terms of $\bv_{_\perp}$, $\delta \rho$, and its derivatives. To lowest (zeroth) order in derivatives, $\delta v_{_\parallel} \approx -{\Gamma_2\over \Gamma_1} \delta\rho$. Inserting this approximate expression for $\delta v_{_\parallel} $ into equation (\ref{v par 1}) everywhere  $\delta v_{_\parallel}$ appears on the right hand side of that equation gives $\delta v_{_\parallel}$ to first order in derivatives:
\begin{widetext}
\begin{eqnarray}
\delta v_{_\parallel}\approx -{\Gamma_2\over \Gamma_1}\left( \delta\rho-{1\over v_0\Gamma_1}\partial_t \delta \rho+{\lambda_4
\partial_{_\parallel} \delta\rho\over\Gamma_2}\right)-{\lambda_2\over\Gamma_1} {\bf\nabla}_{_\perp}\cdot\bv_{_\perp}-{|\bv_{_\perp}|^2\over 2 v_0}~,
\label{v par 2}
\end{eqnarray}
\end{widetext}
where I've defined
\begin{widetext}
\begin{eqnarray}
\lambda_4 \equiv {(\mu_1v_0^2+\sigma_1)\over v_0} - 
{\Gamma_2\over\Gamma_1}\left(\lambda_2+{\gamma_2\over v_0}
\right)= {(\mu_1v_0^2+\sigma_1)\over v_0} - 
{\Gamma_2\over\Gamma_1}\left(\lambda_1+\lambda_2+2\lambda_3
\right)~.
\label{lambda_4 def}
\end{eqnarray}
\end{widetext}
In deriving the second equality in (\ref{lambda_4 def}), I've used the definition (\ref{gamma_2def}) of $\gamma_2$.


 
Inserting (\ref{vexp2}), (\ref{speed}), and 
(\ref{v par 2}) into the equation of motion (\ref{EOM2}) for
$\bv$, and projecting that equation perpendicular to the mean direction of flock motion $\hat{x}_{_\parallel}$ 
gives, neglecting ``irrelevant'' terms:
\begin{widetext}
\begin{eqnarray}
\partial_{t} \bv_{_\perp} + \gamma\partial_{{_\parallel}} 
\bv_{_\perp} &+& \lambda^0_1 \left(\bv_{_\perp} \cdot
{\bf\nabla}_{_\perp}\right) \bv_{_\perp} =-g_1\delta\rho\partial_{{_\parallel}} 
\bv_{_\perp}-g_2\bv_{_\perp}\partial_{{_\parallel}}
\delta\rho -{c_0^2\over\rho_0}{\bf\nabla}_{_\perp}
\delta\rho -g_3{\bf\nabla}_{_\perp}(\delta \rho^2)\nonumber\\&+&
D_B{\bf\nabla}_{_\perp}\left({\bf\nabla}_{_\perp}\cdot\bv_{_\perp}\right)+
D_T\nabla^{2}_{_\perp}\bv_{_\perp} +
D_{{_\parallel}}\partial^{2}_{{_\parallel}}\bv_{_\perp}+\nu_t\partial_t{\bf\nabla}_{_\perp}\delta\rho+\nu_{_\parallel}\partial_{_\parallel}{\bf\nabla}_{_\perp}\delta\rho+\bff_{\perp} \,,\nonumber\\
\label{vEOMbroken}
\end{eqnarray}
\end{widetext}
where  I've defined
\begin{eqnarray}
D_B\equiv D_1+{2v_0\lambda_3^0\lambda_2^0\over\Gamma_1}~, 
\label{DBeff def}
\end{eqnarray}
\begin{eqnarray}
D_{{_\parallel}}\equiv D_{T}+D_{2}v_0^2~,
\label{Dpar def}
\end{eqnarray}
\begin{eqnarray}\gamma \equiv\lambda^0_1v_0~ , 
\label{gamma_1 def}
\end{eqnarray}\begin{eqnarray}
g_{1}\equiv v_0\left({\partial\lambda_1 \over \partial \rho}\right)_0 -{\Gamma_2\lambda_1^0\over\Gamma_1}~, 
\label{g_1 def}
\end{eqnarray}
\begin{eqnarray}
g_2 \equiv{\Gamma_2\gamma_2^0\over\Gamma_1v_0}-{\sigma_1\over v_0}~, 
\label{g_2  def}
\end{eqnarray}
\begin{eqnarray}
g_3 \equiv\sigma_2+\left({\Gamma_2\over\Gamma_1}\right)^2\lambda^0_3-\left({\partial\lambda_3
\over \partial \rho}\right)_0 {\Gamma_2v_0\over\Gamma_1}~,
\label{g_3  def}
\end{eqnarray}

\begin{eqnarray}
 c_0^2 \equiv \rho_0\sigma_1 - {2\rho_0v_0\lambda_3^0\Gamma_2\over \Gamma_1}~, 
 \label{c0def}
 \end{eqnarray}
 
\begin{eqnarray}
\nu_{t}
\equiv -{2\Gamma_2\lambda_3^0\over \Gamma_1^2}~,
\label{nu_t def}
\end{eqnarray}
and
\begin{eqnarray}
\nu_{_\parallel}\equiv {2v_0\lambda_3^0\lambda_4^0\over \Gamma_1}+{\Gamma_2D_1\over\Gamma_1}~.
\label{nu par def}
\end{eqnarray}


 
Using  (\ref{vexp2}) and (\ref{speed}) in the equation of motion (\ref{conservation}) for $\rho$ gives, again neglecting irrelevant terms:
\begin{widetext}
\begin{eqnarray}
\partial_t\delta
\rho +\rho_o{\bf\nabla}_{_\perp}\cdot\bv_{_\perp}
+w_1{\bf\nabla}_{_\perp}\cdot(\bv_{_\perp}\delta\rho)+v_2
\partial_{{_\parallel}}\delta
\rho &=&D_{\rho{_\parallel}}\partial^2_{_\parallel}\delta\rho+D_{\rho{_\perp}}\nabla^2_{_\perp}\delta\rho+D_{\rho v} \partial_{{_\parallel}}
\left({\bf\nabla}_{_\perp} \cdot \bv_{_\perp}\right)\nonumber\\&&+\phi\partial_t\partial_{_\parallel}\delta\rho
+w_2\partial_{_\parallel}(\delta\rho^2)+w_3\partial_{_\parallel}(|
\bv_{_\perp}|^2)~, \nonumber \\
\label{cons broken}
\end{eqnarray}
\end{widetext}
where I've defined:
\begin{eqnarray}
v_2\equiv v_0 - {\rho_0\Gamma_2\over \Gamma_1},~
\label{v_2 def1}
\end{eqnarray}
\begin{eqnarray}
\phi\equiv {\Gamma_2\rho_0\over v_0\Gamma_1^2},~
\label{phi def}
\end{eqnarray}
\begin{eqnarray}
w_2 \equiv {\Gamma_2\over \Gamma_1}~ ,
\label{w_2 def}
\end{eqnarray}
\begin{eqnarray}
w_3 \equiv {\rho_0\over 2v_0}~ ,
\label{w_3 def}
\end{eqnarray}
\begin{eqnarray}
D_{\rho{_\parallel}}&\equiv&{\rho_0\lambda^0_4\over\Gamma_1}\nonumber\\
&= &{\rho_0\over \Gamma_1}\left({(\mu_1v_0^2+\sigma_1)\over v_0}-{\Gamma_2\over \Gamma_1} \left(\lambda^0_1 + \lambda^0_2 + 2\lambda^0_3 \right) \right),
\nonumber \\
\label{Drhopardef}
\end{eqnarray}
and, last but by no means least,
\begin{eqnarray}
D_{\rho v} \equiv {\lambda^0_2\rho_o\over \Gamma_1}, 
\label{D_rho v def}
\end{eqnarray}

The parameter $D_{\rho{_\perp}}$ is actually zero at this point in the calculation, but I've included it in equation (\ref{cons broken}) anyway, because it is generated by the nonlinear terms under the Renormalization Group, as I'll discuss in section (\ref{Sec: nonlinear}). Likewise, the parameter $w_1=1$, but I've also included it for convenience in discussing the renormalization group in section (\ref{Sec: nonlinear}).

I will henceforth focus my attention on the fluid, orientationally ordered state, in which all of the diffusion constants $
D_{\rho{_\parallel}}$, $D_{\rho{_\perp}}$, $
D_{\rho v}$, $
D_B$, $D_{{_\parallel}}$, and $D_T$ are positive. I'll take them all to have their steady state values $D_T^0$ etc. at $|\bv|=v_0$ and $\rho=\rho_0$, since fluctuations away from that can be shown to be irrelevant.

\subsection{\label{Sec: linear}Linearized Theory}

Expanding
(\ref{vEOMbroken}) and (\ref{cons broken})  to linear order in the small fluctuations $\bv_{_\perp}$ and $\delta\rho$ gives:
\begin{widetext}
\begin{eqnarray}
\partial_{t} \bv_{_\perp} + \gamma\partial_{{_\parallel}} 
\bv_{_\perp} & = -{c_0^2\over\rho_0}{\bf\nabla}_{_\perp}
\delta\rho +
D_B{\bf\nabla}_{_\perp}\left({\bf\nabla}_{_\perp}\cdot\bv_{_\perp}\right)+
D_T\nabla^{2}_{_\perp}\bv_{_\perp} +
D_{{_\parallel}}\partial^{2}_{{_\parallel}}\bv_{_\perp}+\nu_t\partial_t{\bf\nabla}_{_\perp}\delta\rho+\nu_{_\parallel}\partial_{_\parallel}{\bf\nabla}_{_\perp}\delta\rho+\bff_{\perp} \,,\nonumber\\
\label{vEOMbroken lin}
\end{eqnarray}
\end{widetext}
and
\begin{widetext}
\begin{eqnarray}
\partial_t\delta
\rho +\rho_o{\bf\nabla}_{_\perp}\cdot\bv_{_\perp}
+v_2
\partial_{{_\parallel}}\delta
\rho =D_{\rho{_\parallel}}\partial^2_{_\parallel}\delta\rho+D_{\rho{_\perp}}\nabla^2_{_\perp}\delta\rho+D_{\rho v} \partial_{{_\parallel}}
\left({\bf\nabla}_{_\perp} \cdot \bv_{_\perp}\right)+\phi\partial_t\partial_{_\parallel}\delta\rho
~.
\label{cons broken lin}
\end{eqnarray}
\end{widetext}

These equations can now readily be solved for the mode structure and correlations by Fourier transforming in space and time; this  gives
\begin{widetext}
\begin{eqnarray}
\left[-i(\omega-\gamma q_{_\parallel}) + \Gamma_L(\bq)\right] v_L+\left[{ic_0^2\over\rho_0}q_{_\perp}-
\nu_t q_{_\perp}\omega-\nu_{_\parallel} q_{_\perp} q_{_\parallel}\right] \delta\rho = f_L
\label{vLEOMFT}
\end{eqnarray}
\end{widetext}
\begin{eqnarray}
\left[-i(\omega-\gamma q_{_\parallel}) + \Gamma_T(\bq)\right] \bv_T = \bff_T
\label{vTEOMFT}
\end{eqnarray}
\begin{widetext}
\begin{eqnarray}
\left[i\rho_0 q_{_\perp} + D_{\rho v}q_{_\perp} q_{_\parallel}\right] v_L+\left[-i(\omega-v_2 q_{_\parallel}) + \Gamma_\rho(\bq)-\phi q_{_\parallel}\omega\right] \delta\rho = 0
\label{rhoEOMFT}
\end{eqnarray}
\end{widetext}
and
where I've defined
the wavevector 
dependent  longitudinal, transverse, and $\rho$ dampings $\Gamma_{L,T, \rho}$:
\begin{eqnarray}
\Gamma_L\left(\bq\right)  = D_L q_{_\perp}^2 + D_{{_\parallel}}q^2_{{_\parallel}} \, ,
\label{GLdef}
\end{eqnarray}
\begin{eqnarray}
\Gamma_T\left(\bq\right)  = D_T q_{_\perp}^2 + 
D_{{_\parallel}}q^2_{{_\parallel}} \, ,
\label{GTdef}
\end{eqnarray}
\begin{eqnarray}\Gamma_{\rho}
\left(\bq\right) = D_{\rho{_\parallel}} q^2_{{_\parallel}} +D_{\rho{_\perp}}q^2_{{_\perp}}\, ,
\label{Grhodef}
\end{eqnarray}
with $D_L
\equiv D_B+D_T$.
I've also separated the velocity $\bv_{_\perp}$ and the noise $\bff_\perp$ into components along and perpendicular to the projection $\bq_{_\perp}$
of $\bv$  perpendicular to $<\bv>$ via 
\begin{eqnarray}
v_L\equiv \bv_{_\perp}\cdot\bq_{_\perp}/q_{_\perp}
~,~\bv_T\equiv \bv_{_\perp}- v_L {\bq_{_\perp}\over q_{_\perp}}~,
\label{vLTdef}
\end{eqnarray}
with $f_L$ and $\bff_T$ obtained from $\bff$ in the same way.

These equations differ from the  corresponding equations considered in \cite{TT1,TT2,TT3,TT4} only in the $\nu_{t,{_\parallel}}$ terms in  (\ref{vLEOMFT}),  and the $D_{\rho{_\parallel}}$ and $D_{\rho v}$ terms in
(\ref{rhoEOMFT}). These prove to lead only to minor changes in the propagation direction dependence, but not the scaling with wavelength, of the damping of the sound modes found in \cite{TT1,TT2,TT3,TT4}, as I will now demonstrate.

\begin{figure}
 \includegraphics[width=1.0\textwidth]{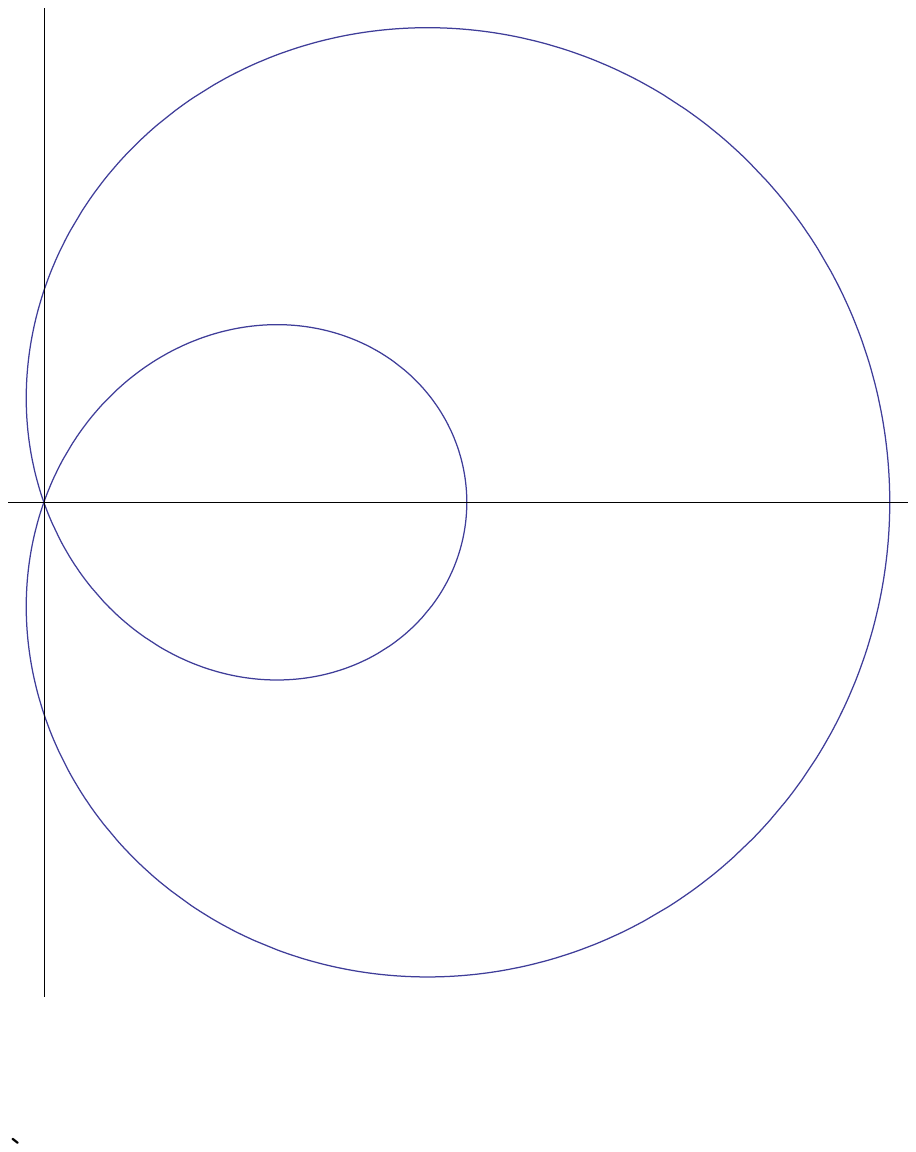}
\caption{\label{sound}Polar plot of the direction-dependent sound speeds $c_{\pm} \left(\theta_{\bq}\right)$, with the horizontal axis along the direction of mean flock motion.}
\end{figure}

I begin by determining the eigenfrequencies of the system, defined in the usual way 
as the complex, wavevector dependent  frequencies $\omega(\bq)$ at which the Fourier transformed hydrodynamic equations (\ref{vLEOMFT}),  (\ref{vTEOMFT}), and  (\ref{rhoEOMFT}) admit non-zero solutions for 
$\bv_T$, $\delta \rho$, and $v_L$ when the noise $\bff$ is set to zero. Note that $\bv_T$ is decoupled from $v_L$ and $\rho$; this implies a pair of ``longitudinal" eigenmodes involving just the longitudinal velocity $v_L$ and $\rho$, and an additional $d-2$ ``transverse" modes associated with the transverse velocity $\bv_T$. The longitudinal modes are closely analogous to ordinary sound waves in a simple fluid\cite{Forster}, while the transverse modes are the analog of the diffusive shear modes in such a fluid.

In the hydrodynamic  limit (i.e., when  wavenumber $q\rightarrow 0$), the longitudinal eigenfrequencies become a pair of underdamped, propagating modes with complex eigenfrequencies
\begin{eqnarray}
\omega_{\pm}(\bq)=c_{\pm}(\theta_{\bq}) q - i\epsilon_{\pm}({\bq})~,
\end{eqnarray}
where the direction-dependent sound speeds $c_{\pm}\left(\theta_{\bq}\right)$ are given by exactly the same expression as found in previous work\cite{TT1,TT2,TT3,TT4}:
\begin{eqnarray}
c_{\pm}\left(\theta_{\bq} \right) =\left({\gamma + v_2 \over 2}\right)\cos\left(\theta_{\bq}\right)\pm c_2\left(\theta_{\bq} \right) ,\label{cplusminus}\end{eqnarray}where I've defined\begin{eqnarray}c_2\left(\theta_{\bq} \right) \equiv\sqrt{{\left(\gamma -v_2\right)^2 \cos^2\left(\theta_{\bq}\right) \over 4} + c_0^2\sin^2 \left(\theta_{\bq} \right)} \quad , \nonumber \\ \label{c2}\end{eqnarray}where $\theta_{\bq}$ is the angle between $\bq$ and the direction of flock motion (i.\ e., the $x_{_\parallel}$ axis).

A polar plot of this highly anisotropic sound speed is given in figure (\ref{sound}). 

This prediction for the anisotropy of the sound speeds in flocks has recently been confirmed quantitatively in experiments on synthetic flockers (specifically, Quinke rotators) \cite{Bartolo2}.

As mentioned earlier, the wavevector dependent dampings $\epsilon_{\pm}(\bq)$ of these propagating sound modes {\it are} altered slightly from the form found in \cite{TT1,TT2,TT3,TT4}. They remain of $O(q^2)$, as found in previous work, but with a slightly modified dependence on propagation direction $\hat{q}$. More precisely, they are given by:
\begin{widetext}
\begin{eqnarray}
\epsilon_{\pm}\equiv{{\rm{Hideous \, Numerator}}\over(2c_{\pm}(\theta_{\bq})-(v_2+\gamma){\cos}(\theta_{\bq}))}
\label{damping 1}
\end{eqnarray}
\end{widetext}
with
\begin{widetext}
\begin{eqnarray}
\rm{Hideous \, Numerator}&\equiv&(\Gamma_L(\bq) +
\Gamma_\rho(\bq)
-\phi c_{\pm}(\theta_{\bq})\cos(\theta_{\bq})q^2) c_{\pm}(\theta_{\bq})-v_2\Gamma_L(\bq)\cos(\theta_{\bq})\nonumber\\
&-&\gamma(\Gamma_\rho(\bq)-\phi c_{\pm}(\theta_{\bq})\cos(\theta_{\bq})q^2)\cos(\theta_{\bq})+{c_0^2\over \rho_0}D_{\rho v}{q_{_\parallel} q_{_\perp}^2\over q}\nonumber\\&-&\rho_0 q_{_\perp}^2(\nu_t c_{\pm}(\theta_{\bq})+\nu_{_\parallel} \cos(\theta_{\bq}))~,\nonumber\\
\label{damping 2}
\end{eqnarray}
\end{widetext}
where I remind the reader that the wavevector dependent dampings $\Gamma_{L,\rho}$ are $O(q^2)$, and defined earlier
in equations (\ref{GLdef}, \ref{Grhodef}). Thus, the ``hideous numerator", while indeed hideous in its angular dependence, is nonetheless simple in its scaling with the {\it magnitude} $q$ of $\bq$: it scales like $q^2$. This implies that the dampings $\epsilon_\pm\propto q^2$ as well.

The transverse modes have the far simpler character of simply convected anisotropic diffusion: \begin{eqnarray}
\omega_T(\bq)=\gamma q_\parallel-i\Gamma_T(\bq)
\label{omegaT}
\end{eqnarray}
with the wavevector dependent damping $\Gamma_{T}$ also $O(q^2)$, and defined earlier
in equation (\ref{GTdef}]). This corresponds to simple anisotropic diffusion in a ``pseudo-comoving" frame, by which I mean a frame that moves in the direction of mean flock motion, but with a speed $\gamma$ that differs from the speed $v_0$ of the flock itself.


I now turn to the correlation functions in this linearized approximation. These are easily obtained by first solving the linear algebraic equations (\ref{vLEOMFT}),  (\ref{vTEOMFT}), and  (\ref{rhoEOMFT})
for the fields $v_L(\bq, \omega)$, $\bv_T(\bq, \omega)$, and $\rho(\bq, \omega)$
in terms of the noises $f_L(\bq, \omega)$, and $\bff_T(\bq, \omega)$. These solutions are, of course, linear in those noises. Hence,  by correlating these solutions pairwise, one can obtain any two field correlation function in terms of the correlations (\ref{white noise}) of $\bff$. The resulting correlation function for the velocity  is:
\begin{widetext}
\begin{eqnarray}
C_{ij}\left(\bq,\omega\right) \equiv\left<v_{\perp i}\left(\bq,\omega\right) v_{\perp j}\left(-\bq,-\omega\right)\right>&=&{\Delta\left(\omega - v_2q_{{_\parallel}}\right)^2L^\perp_{ij}\over\left[(\omega-c_+\left(\theta_{\bq}\right)q)^2+\epsilon_+^2(\bq)\right] \left[(\omega-c_-\left(\theta_{\bq}\right)q)^2+\epsilon_-^2(\bq)\right] }\nonumber\\&+&{\Delta P^\perp_{ij}\over\left[(\omega-\gamma q_{_\parallel})^2+\Gamma_T^2(\bq)\right] }~.
\label{CL}
\end{eqnarray}
\end{widetext}
where I've defined the  longitudinal (L) and transverse (T) projection operators in the $\perp$ plane
\begin{eqnarray}
L^\perp_{ij}(\hat{q})\equiv{q_{\perp i}q_{\perp j}\over q_{_\perp}^2}~,~P^\perp_{ij}(\hat{q})\equiv \delta^\perp_{ij}-L^\perp_{ij}(\hat{q})~,
\label{proj}
\end{eqnarray}
where $\delta^\perp_{ij}$ is a Kronecker delta in the $\perp$ plane (i.e., it is equal to the usual Kronecker delta if $i\ne{\parallel}\ne j$, and zero otherwise). These operators project any vector first into the $\perp$ plane, and then either along (L) or orthogonal to (P) $\bq_{_\perp}$ within the $\perp$ plane.

The first term in equation (\ref{CL}) comes from the ``longitudinal" component $v_L$ 
while the second comes from the $d-2$ ``transverse" components of $\bv_{_\perp}$. Clearly, in $d=2$, only the longitudinal component is present; the second (transverse) term in (\ref{CL}) vanishes in $d=2$.

The density autocorrelations obtained by the procedure described above are given, to leading order in wavevector and frequency,  by:
\begin{widetext}
\begin{eqnarray}
C_{\rho\rho}\left(\bq,\omega\right) \equiv\left<\rho\left(\bq,\omega\right) \rho\left(-\bq,-\omega\right)\right>&=&{\rho_0 q_{_\perp}^2\Delta\over\left[(\omega-c_+\left(\theta_{\bq}\right)q)^2+\epsilon_+^2(\bq)\right] \left[(\omega-c_-\left(\theta_{\bq}\right)q)^2+\epsilon_-^2(\bq)\right] }~.\nonumber\\
\label{Crho}
\end{eqnarray}
\end{widetext}

Both the velocity correlations (\ref{CL}) and the  density correlations (\ref{Crho}) have the same form, and 
the same scaling with frequency  and wavevector, as those reported in earlier work
\cite{TT1,TT2,TT3,TT4}. The only change from those earlier results is the slightly modified 
form (\ref{damping 1}, \ref{damping 2}) of the sound dampings which appear in  (\ref{CL}) and  (\ref{Crho}).

The same statement is true of the equal-time correlations of $\bv$ and $\rho$, which can be obtained in the usual way by integrating the spatiotemporally Fourier transformed correlations  (\ref{CL}) and  (\ref{Crho}) over all frequency $\omega$. These equal time correlations are important, because they determine the size of the velocity and density fluctuations. The size of the velocity fluctuations determines whether or not long ranged order can exist in these systems, while the size of the density fluctuations determines the presence or absence of giant number fluctuations\cite{TT4, actnemsub, Chate+Giann, act nem}.

Integrating (\ref{CL}) over all $\omega$ and tracing over the Cartesian components $i,j$ gives the equal-time correlation of $\bv$:
\begin{widetext}
\begin{eqnarray}
\left<|\bv_{_\perp}\left(\bq,t\right) |^2\right>&=&{1\over 2}\left({\Delta\left(c_+\left(\theta_{\bq}\right) - v_2\cos
\left(\theta_{\bq}\right)\right)^2\over\epsilon_+\left({\bq}\right)[c_+\left(\theta_{\bq}\right)-c_-\left(\theta_{\bq}
\right)]^2}+{\Delta \left(c_-\left(\theta_{\bq}\right) - v_2\cos\left(\theta_{\bq}\right)\right)^2\over
\epsilon_-\left({\bq}\right)[c_+\left(\theta_{\bq}\right)-c_-\left(\theta_{\bq}\right)]^2}+{(d-2)\Delta\over\Gamma_T(\bq)}\right)~.\nonumber\\
\label{vET}
\end{eqnarray}
\end{widetext}
Note that these scale like $1/q^2$ for all directions of wavevector $\bq$. This scaling is precisely the same as that found in  the linearized theory of\cite{TT1,TT2,TT3,TT4}; only the precise form of the dependence on the direction of $\bq$ is slightly changed by the presence of the new linear terms $\nu_t$, $\nu_\parallel$, and $\phi$ that I've found here that were missed in the treatment of\cite{TT1,TT2,TT3,TT4}.

This $1/q^2$ scaling of $\bv_\perp$ fluctuations with $q$ in Fourier space implies that the real space fluctuations 
\begin{eqnarray}
\left<|\bv_{_\perp}\left(\br,t\right) |^2\right>&=&\int {d^dq\over(2\pi)^d}\left<|\bv_{_\perp}\left(\bq,t\right) |^2\right>
\label{vreal}
\end{eqnarray}
diverge in the infra-red ($q\rightarrow0$ or system size $L\rightarrow\infty$) limit in all  spatial dimensions $d\le2$. This in turn implies that long-ranged order (i.e., the existence of a non-zero $\left<\bv_{_\perp}\left(\br,t\right)\right>$) is not possible in $d=2$, {\it according to the linearized theory}.

This result, which is simply the Mermin-Wagner\cite{MW} theorem, is actually overturned by non-linear effects, which stabilize the long-ranged order in $d=2$ (i.e., make the existence of a non-zero $\left<\bv\left(\br,t\right)\right>$ possible), as first noted by\cite{TT1,TT2,TT3,TT4}. I'll show in subsection (\ref{Sec: nonlinear}) that non-linear effects still stabilize long-ranged order in this way even when the additional nonlinearities I've found here, which  were missed in \cite{TT1,TT2,TT3,TT4}, are included.

The equal time density autocorrelations can likewise be obtained by integrating equation (\ref{Crho}) over frequency $\omega$; this gives
\begin{widetext}
\begin{eqnarray}
\left<|\delta\rho\left(\bq,t\right) |^2\right>&=&{1\over 2}\left({\Delta\rho_0q_\perp^2\over[c_+\left(\theta_{\bq}\right)-c_-\left(\theta_{\bq}
\right)]^2 q^2}\right)
\left({1\over\epsilon_+\left({\bq}\right)}+{1\over\epsilon_-\left({\bq}\right)}\right)
\label{rhoET}
\end{eqnarray}
\end{widetext}
This also scale like $1/q^2$  for all directions of $\bq$. This divergence implies ``Giant Number Fluctuations"\cite{TT4, actnemsub, Chate+Giann, act nem}: the RMS fluctuations 
$\sqrt{\left<\delta N ^2\right>}$ 
of the number of particles within a large region of the system scale like the mean number of particles $\left<N \right>$ faster than $\sqrt{\left<N \right>}$; specifically, $\sqrt{\left<\delta N ^2\right>}\propto\left<N \right>^{\phi(d)}$, with $\phi(d)=1/2+1/d$ in spatial dimension $d$. Note that this means in particular that
$\sqrt{\left<\delta N ^2\right>}\propto\left<N \right>$ in $d=2$.

Again, I emphasize that this is the prediction of the linearized theory. It once again coincides with the results of the linearized treatment of \cite{TT1,TT2,TT3,TT4}.

Both the prediction  that long ranged orientational order is destroyed in $d=2$, and the value   $\phi(d)=1/2+1/d$ of the exponent $\phi(d)$ for $d<4$ prove, when non-linear effects are taken into account, to be incorrect, as first noted by \cite{TT1}. I now turn to the treatment of those nonlinear effects.

\subsection{\label{Sec: nonlinear}Non-linear Effects}

We have seen that the linearized theory does not explain the mystery that motivated my original interest in this problem: the persistence of long-ranged order in flocks even in $d=2$. Fortunately, it turns out that the non-linearities that we ignored in the previous section in fact  completely change the scaling behavior of these systems at long distances, as first noted by \cite{TT1,TT2,TT3,TT4}. In this section, I'll deal with those non-linearities. While a few of the precise quantitative conclusions of \cite{TT1,TT2,TT3,TT4} prove to be less certain than Yuhai and I originally thought, the essential conclusions that 

\noindent 1) non-linearities radically change the scaling behavior of these systems for all $d\le4$, and

\noindent 2) these changes in scaling stabilize long-ranged order in $d=2$,

\noindent remain valid.


Equally noteworthy are the non-linear terms that are missing from (\ref{vEOMbroken}) and (\ref{cons broken}): all nonlinearities arising from the anisotropic pressure $P_2$ and the $\lambda_2$ nonlinearity drop out of (\ref{vEOMbroken}) and (\ref{cons broken}). This in particular has the very important consequence of saving the Mermin-Wagner theorem. This is because  the $\lambda_2$  term is allowed even in equilibrium systems \cite{Marchetti}. The incorrect treatment in \cite{TT1,TT2,TT3,TT4} suggested that this term {\it by itself} could stabilize long-range order in $d=2$. Given that this term is allowed in equilibrium, this would imply that the Mermin-Wagner theorem would fail for such an equilibrium system. The correct treatment I've done here shows that this is not the case: the  $\lambda_2$ term by itself  cannot stabilize long ranged order in $d=2$, since the non-linearities associated with it drop out of the long-wavelength description of the ordered phase.

Returning now to the non-linearities in 
(\ref{vEOMbroken}) and (\ref{cons broken}) that were missed by \cite{TT1,TT2,TT3,TT4}, I will now show that {\it all} of them become relevant, in the renormalization group (RG) sense\cite{Ma's book}, for spatial dimensions  $d\le 4$.

To assess the effect of the new non-linear terms I've found here, I'll analyze equations 
(\ref{vEOMbroken}])and (\ref{cons broken}) 
 using the dynamical Renormalization Group(RG)\cite{FNS}. 

The dynamical RG starts by averaging the equations of motion over the short-wavelength fluctuations: i.e.,   those with support in the ``shell" of Fourier space $b^{-1} \Lambda \le |\bq| \le \Lambda$, where $\Lambda$ is an ``ultra-violet cutoff", and $b$ is an arbitrary rescaling factor. Then, one  rescales lengths, time,  $\delta\rho$ and $\bv_{_\perp}$ in equations  (\ref{vEOMbroken}) and (\ref{cons broken}) according to 
$\bv_{_\perp} = b^\chi \bv_{_\perp}^{\,\prime}$, $\delta\rho= b^\chi \delta\rho^{\,\prime}$, $\br_\perp=b\br_\perp^{\,\prime}$, $r'_
{_\parallel}=b^{\zeta}(r'_{_\parallel})'$, and $t = b^zt'$ to restore the ultra-violet cutoff  to $\Lambda$ \cite{chirho}.  
This leads to a new pair of equations of motion of the same form as (\ref{vEOMbroken}) and (\ref{cons broken}) , but with 
``renormalized" values (denoted by primes below) of the parameters given by:
\begin{eqnarray} 
D'_{B, T} = b^{z-2}(D_{B,T}+  {\rm 
graphs} )~~, 
\label{rescale Dperp}    
\end{eqnarray}
\begin{eqnarray} 
D'_{{_\parallel},\rho\parallel}=b^{z-2\zeta}(D_{{_\parallel},\rho\parallel} + {\rm 
graphs} )~~,
\label{rescale Dpar}   
\end{eqnarray}
\begin{eqnarray} 
\Delta' = b^{z-\zeta-2\chi+1-d}(\Delta+{\rm 
graphs} )~,
\label{rescale Delta}    
\end{eqnarray}
\begin{eqnarray} 
(\lambda^0_{1})'= b^{z+\chi-1}(\lambda^0_1+{\rm 
graphs} )~~
,
\label{rescale lambda} 
\end{eqnarray} 
\begin{eqnarray} 
g'_{1,2}= b^{z+\chi-\zeta}(g_{1,2}+{\rm 
graphs} )~~
,
\label{rescale g12} 
\end{eqnarray} 

\begin{eqnarray} 
g'_{3}= b^{z+\chi-1}(g_{3}+{\rm 
graphs} )~~
,
\label{rescale g3} 
\end{eqnarray} 

\begin{eqnarray} 
\phi'= b^{z+\chi-1}(\phi+{\rm 
graphs} )~~
,
\label{rescale phi} 
\end{eqnarray} 

\begin{eqnarray} 
w'_{1,2,3}= b^{z+\chi-\zeta}(w_{1,2,3}+{\rm 
graphs} )~~
,
\label{rescale w} 
\end{eqnarray} 
where ``graphs" denotes contributions from integrating out the short wavelength degrees of freedom.

I have focused on the particular linear parameters $D_{B,T,{_\parallel},\rho\parallel}$ and $\Delta$ since, as is clear from equations (\ref{vET})  and  (\ref{rhoET}), they determine 
the size of the fluctuations in the linearized theory.

One proceeds by seeking fixed points of these recursion relations.
One simple fixed point is the linear fixed point, at which all of the non-linear coefficients 
$\lambda_1^0$, $g_{1,2,3}$, and $w_{1,2,3}$ are zero. At such a fixed point, the graphical corrections (denoted "graphs"
in equations  (\ref{rescale Dperp}) - (\ref{rescale w})) vanish, since, without nonlinearities, Fourier modes at different wavevectors and frequencies do not interact. It is then straightforward to determine from equations  (\ref{rescale Dperp}) - (\ref{rescale w})) the values of the rescaling
exponents $z$, $\zeta$, and  $\chi$
that will keep $D_{B,T,{_\parallel},\rho\parallel}$ and $\Delta$ (and, hence, the size of the fluctuations) fixed: simply those that make the exponents in  (\ref{rescale Dperp}) - (\ref{rescale w})) vanish. 
That is, we must chose
\begin{eqnarray} 
z-2=0~~\rm{(linear}~\rm{  fixed}~\rm{  point)}
\label{fix Dperp}    
\end{eqnarray}
to keep $D_B$ and $D_T$ fixed, 
\begin{eqnarray} 
z-2\zeta=0~~\rm{(linear}~\rm{  fixed}~\rm{  point)},
\label{fix Dpar}    
\end{eqnarray}
to keep $D_{{_\parallel}}$ and $D_{\rho\parallel}$ fixed, and 
\begin{eqnarray} 
z-\zeta-2\chi+1-d=0~~\rm{(linear}~\rm{  fixed}~\rm{  point)}, 
\label{fix Delta}    
\end{eqnarray}
to keep $\Delta$ fixed under the RG. The solutions to these three conditions (\ref{fix Dperp})-(\ref{fix Delta})
are trivially found to be: 
\begin{eqnarray} 
z=2~~ \rm{(linear}~\rm{  fixed}~\rm{  point)}, 
\label{zlin}    
\end{eqnarray}
\begin{eqnarray} 
\zeta=1~~\rm{(linear}~\rm{  fixed}~\rm{  point)}, 
\label{zetalin}    
\end{eqnarray}
and
\begin{eqnarray} 
\chi=(2-d)/2~~\rm{(linear}~\rm{  fixed}~\rm{  point)} . 
\label{chilin}    
\end{eqnarray}

Let's now consider the stability this linear fixed point  against the effect of the non-linear terms $\lambda_1^0$, $g_{1,2,3}$, and $w_{1,2,3}$. Because, as mentioned earlier,  I have chosen the rescaling exponents so as to keep the magnitude of the fluctuations the same on all length scales,  a given non-linearity has important effects at long distances if it grows upon renormalization with this choice (\ref{zlin})-(\ref{chilin}) of  the rescaling
exponents $z$, $\zeta$, and  $\chi$ provided that it grows upon renormalization; contrariwise, if it gets smaller upon renormalization with 
 this choice of  the rescaling
exponents, it is unimportant at long distances\cite{couplings}. 
Using the exponents (\ref{zlin})-(\ref{chilin}) in the recursion relations (\ref{rescale lambda}),  (\ref{rescale g12}), (\ref{rescale g3}),(\ref{rescale phi}),and (\ref{rescale w}), and ignoring the graphical corrections, which are higher than linear order in $\lambda_1^0$, $g_{1,2,3}$, and $w_{1,2,3}$,  I find that all seven of these non-linearities have identical renormalization group eigenvalues of $(4-d)/2$ at the linearized fixed point; that is:
\begin{eqnarray} 
(\lambda^0_{1})'= b^{4-d\over 2}\lambda^0_1~~
,
\label{rescale lambda} 
\end{eqnarray} 
\begin{eqnarray} 
g'_{1,2,3}= b^{4-d\over 2}g_{1,2,3}~~
,
\label{rescale g12} 
\end{eqnarray} 

\begin{eqnarray} 
w'_{1,2,3}= b^{4-d\over 2}w_{1,2,3}~~.
\end{eqnarray}

Thus, for $d>4$, all of the nonlinearities flow to zero, and so become unimportant, at long length and time scales. Hence, the linearized theory is correct at  long length and time scales, for $d>4$. For $d<4$, however, all of these nonlinearities grow, and the linear theory breaks down at sufficiently long length and time scales.

Both this analysis, and its conclusion that non-linear effects invalidate the linear theory for $d<4$, are {\it almost} identical  to those of \cite{TT1,TT2,TT3,TT4}. However, whereas they found only four non-linearities ($\lambda_{1,2}$, $w_1$,   and $g_3$ in the notation I'm using here) that became relevant as $d$ is decreased below $d=4$, I find seven such nonlinearities.  More importantly, the vector structure of some of the new nonlinearities differs from that of those studied in \cite{TT1,TT2,TT3,TT4} $\,\ $ in crucial ways. In particular,  all of 
the nonlinearities considered in \cite{TT1,TT2,TT3,TT4} $\,\ $ could, in $d=2$, be written as total $\perp$ derivatives. This implies that these nonlinearities can only 
renormalize terms which themselves involved $\perp$-derivatives (i.e., $D_{B,T}$); hence, all of the terms that did {\it not} involve $\perp$-
derivatives (i.e., $D_{_\parallel,\rho\parallel}, \Delta$) were incorrectly argued in \cite{TT1,TT2,TT3,TT4} $\,\ $  to get no graphical corrections. This lead to the incorrect conclusion that, in order to obtain a fixed point, one had to choose  the rescaling
exponents $z$, $\zeta$, and  $\chi$ to make the exponents in (\ref{rescale Dpar}) and (\ref{rescale Delta}) vanish; i.e.,  that in $d=2$, 
\begin{eqnarray} 
z-2\zeta=0,
z-\zeta-2\chi+1-d=z-\zeta-2\chi-1=0. \nonumber\\ 
\label{harm exp 3}    
\end{eqnarray}

The earlier work of \cite{TT1,TT2,TT3,TT4} went on to incorrectly argue that  there were no graphical corrections 
$\lambda_1^0$ either, because the equations of motion  (\ref{vEOMbroken}) and (\ref{cons broken})
have, in $d=2$ and in the absence of the extra relevant nonlinearities $g_{1,2}$ and $w_{1,2,3}$ found here,  an exact 
``pseudo-Galilean invariance" symmetry\cite{pseudo}: they
remain unchanged by 
a pseudo-Galilean 
the transformation:
\begin{eqnarray} 
\br_\perp \to \br_\perp-\lambda_1 \bv_1 t~~~,~~
\bv_{_\perp} \to \bv_{_\perp} + \bv_1~~~
,\label{Gal} 
\end{eqnarray} 
 for arbitrary constant vector $\bv_1\perp\hat{x}_{_\parallel}$. 
Note that if $\lambda=1$, this reduces to the familiar Galilean invariance in the $x$-direction.
Since such an exact symmetry  must continue to hold upon renormalization, with the {\it same} value of  $\lambda_1$,  $\lambda_1$ cannot be graphically renormalized in the absence of the  extra relevant nonlinearities $g_{1,2}$ and $w_{1,2,3}$ found here.
Requiring that $\lambda_1'=\lambda_1$ in 
(\ref{rescale lambda}),
and  setting ${\rm 
graphs} =0$, implies 
that
\begin{eqnarray}
\chi =1 -z
\label{fix lambda}
\end{eqnarray}
 in  $d=2$.
This and (\ref{harm exp 3}) forms  three independent equations for the three unknown
exponents 
$\chi$, $z$, and $\zeta$,  whose solution in $d=2$ is 
\begin{eqnarray} 
z=6/5~~
\label{zNL}    
\end{eqnarray}
\begin{eqnarray} 
\zeta=3/5~~
\label{zetaNL}    
\end{eqnarray}
and
\begin{eqnarray} 
\chi=-1/5~~,
\label{chiNL}    
\end{eqnarray}
which are the exponents purported in \cite{TT1,TT2,TT3,TT4} $\,\ $ to be exact in $d=2$.


The presence of the extra nonlinearities
 $g_{1,2,3}$ and $w_{1,2,3}$ invalidates every essential ingredient of the above argument: these nonlinearities are not total $\perp$-derivatives, so one can {\it not} argue that $D_{_\parallel,\rho\parallel}$ and $ \Delta$  get no graphical corrections. This invalidates the exact scaling relations (\ref{harm exp 3}), and makes it impossible to obtain exact exponents in $d=2$.

I have been unable to come up with alternative arguments that give exact exponents in the presence of these additional terms.

Now, {\it if} these  additional nonlinearities were irrelevant in $d=2$ under a full dynamical RG, then the exact exponents of \cite{TT1,TT2,TT3,TT4} would be correct in $d=2$.

There is a precedent for this (that is, for terms that appear relevant by simple power counting below some critical dimension $d_c$ actually proving to be irrelevant once "graphical corrections" -i.e., nonlinear fluctuation effects - are taken into account). One example of this is the cubic symmetry breaking interaction\cite{Aharony} in the $O(n)$ model, which is relevant by power counting at the Gaussian fixed point for $d<4$, but proves to be irrelevant, for sufficiently small $n$, at the Wilson-Fisher fixed point that actually controls the transition for $d<4$, at least for $\epsilon\equiv4-d$ sufficiently small.

Unfortunately, doing a similar $4-\epsilon$ analysis of the relevance of these new nonlinearities in the flocking problem would tell us nothing about whether or not
these terms are relevant in $d=2$, since $2$ is far below the critical dimension $d_c=4$ of the flocking problem.

Hence, whether or not the exact exponents predicted by \cite{TT1,TT2,TT3,TT4} are correct remains an open question. They could be; numerical experiments\cite{TT2, TT3, TT4, Chate+Giann, Chate1, Chate2} suggest they are, but we really don't know at this point.

Not all of the predictions of \cite{TT1,TT2,TT3,TT4} become questionable in light of the existence of these new nonlinearities, however. In particular, the claim that long ranged orientational order can exist even in $d=2$ is unaffected. I know this because the nonlinear terms clearly make positive contributions to the damping coefficient corrections to the velocity diffusion ``constants" $D_B$ and $D_{T}$ are positive, and that they are 
relevant in the RG sense, which means they must change the scaling of the velocity fluctuations from that predicted by the linearized theory. I  know that they are relevant by the following  proof by contradiction:  if all of the nonlinear effects were irrelevant, then simple power counting would suffice to determine their relevance. But simple power counting says that {\it all} of the nonlinearities are {\it relevant} for $d<d_c=4$, which contradicts the original assumption that they're all irrelevant.
Thus, the nonlinearites {\it must} change the scaling of the velocity fluctuations. Since the effect of the nonlinearities is to renormalize the velocity diffusion ``constants" $D_B$ and $D_T$  upwards, and since this tends to reduce velocity fluctuations, the growth of velocity fluctuations with length scale must be suppressed (more precisely, its {\it scaling} must be suppressed; i.e., it must grow like a smaller power of length scale $L$)
than is predicted by the linearized version of the equations of motion (\ref{vEOMbroken})and (\ref{cons broken}). But those linearized  equations predict \cite{TT1,TT2,TT3,TT4} only 
{\it logarithmic} divergences of velocity fluctuations with length scale. Hence, the real fluctuations, including nonlinear effects, must be smaller than logarithmic by some power of length scale\cite{no log}, which means they must be {\it finite } as  $L\rightarrow \infty$. This boundedness of velocity fluctuations means that long ranged order {\it is} possible in a two-dimensional flock, in contrast to equilibrium systems with continuous symmetries.

Note that  all of the troublesome nonlinearities that make it impossible to determine exact exponents in $d=2$  involve the fluctuation $\delta\rho$ of the density $\rho$. 
Therefore, if these fluctuations could somehow be ``frozen out", it would be possible to determine exact exponents in $d=2$. 

There are a number of types of flocks  in which precisely such a freezing out of density fluctuations occurs. Two classes of such systems, namely,  flocks with birth and death \cite{Malthus} and incompressible systems\cite{cfl} - have been treated elsewhere. In both of these systems, exact scaling exponents can be found in $d=2$.

By simply keeping track of the rescaling done in the dynamical RG, one can derive scaling laws for the velocity correlations. For example, the  correlations of the perpendicular components $\dvp$ of the velocity (which can be measured in both simulations and experiments, the latter by image analysis) are given by
\beq
\langle\dv_{_\perp}(\mathbf 0,0)\cdot\dv_{_\perp}(\bbr,t)\rangle
\sim\left\{
\begin{array}{ll}
r_{_\perp}^{2\chi},&|x-\gamma t|\ll r_{_\perp}^{\zeta}, |t|\ll r_{_\perp}^z\\
|x-\gamma t|^{2\chi\over\zeta},&|x-\gamma t|\gg r_{_\perp}^{\zeta}, |x-\gamma t| \gg  |t|^{1\over 2}\\
|t|^{2\chi\over z},&|t|\gg r_{_\perp}^z, |t|\gg |x-\gamma t|^2
\label{uscale}
\end{array}\,,
\right.
\eeq
where I remind the reader that, in $d=2$, the conjecture described above leads to the  {\it exact} values of the scaling exponents
\beq
\zeta={3\over 5}\sep z={6\over 5} \sep \chi=-{1\over 5} \ ,
\label{exact}
\eeq


These scaling predictions agree extremely well with numerical
simulations \cite{TT2,TT3,TT4}.

\section{20-20 Hindsight handwaving argument}

In this section, 
I will rederive the scaling results we found above by an extension of the ``blob" derivation of the Mermin-Wagner theorem  given in section (II). 

Consider the group of birds that at time $t$ whose velocities will be well-correlated with what some reference bird was doing at $t=0$. This group will be a ``blob" whose center moves along the direction of mean motion of the flock at a speed  $\gamma$.

I'll start by proving by contradiction that, for spatial dimensions $d\le4$, the motion of the flockers implies that the width of this blob can {\it not} scale like the width of the blob of pointers in section II.  Those of you who've been paying attention will note that the critical dimension of $4$ is exactly what we found in the dynamical RG of the previous section.

If we {\it do} assume that this blob grows like the analogous blob for pointers-that is, diffusively-then it will be essentially isotropic, and have width:

\beq
w(t)\propto t^{1/2}
\label{wdif}
\eeq
as illustrated in figure (\ref{blobwrong}).

\begin{figure}
 \includegraphics[width=1.0\textwidth]{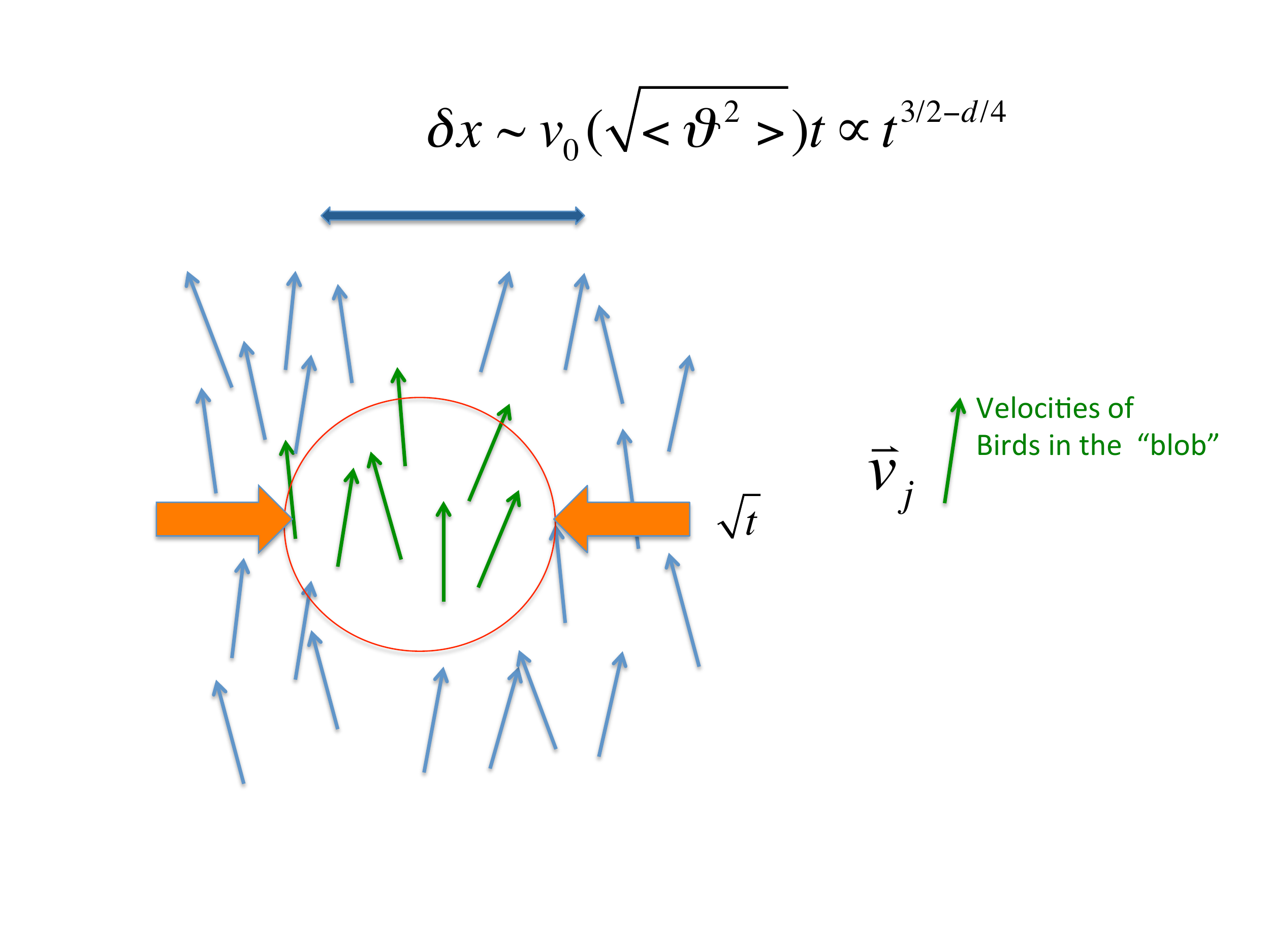}
\caption{\label{blobwrong}Evolution of the ``blob" in the incorrect picture that it grows diffusively in all directions. The lateral wandering of the blob $\delta x$ actually eventually exceeds its diffusive width $w(t)$ for all spatial dimensions $d\le4$.}
\end{figure}

But {\it unlike} the pointers, this blob will now be moving. Indeed, its velocity perpendicular to the mean direction of flock motion will be $\dvp\sim v_0\theta$, where $\theta$ is the deviation of the mean direction of the blob from that of the flock. This means that the blob will wander laterally relative to the rest of the flock.

How far will it  wander laterally  in time $t$?
Roughly
\beq
\delta x\sim\sqrt{<v_\perp^2>}t\sim v_0\sqrt{<\theta^2>}t \,.
\label{delx1}
\eeq
Note that this wandering is purely lateral: fluctuations in the velocity {\it along} the mean direction of flock motion go like $v_0[1-\cos(\theta)]\propto\theta^2$ for small $\theta$, and, so, are much smaller than the lateral velocity fluctuations $\dvp$.

If we now assume (and remember, we're going to be showing that this assumption is actually self-contradictory for $d\le4$), that the fluctuations of $\theta$ of this blob of flockers scale just as those of the blob of pointers in section II, we have: 
\beq
\sqrt{<\theta^2>}\propto r^{1-d/2}\propto t^{1/2-d/4}\,.
\label{deltheta}
\eeq
Using this in (\ref{delx1}), I get 
\beq
\delta x\propto t^{3/2-d/4}
\label{delx2}
\eeq

Comparing this to the width $w(t)$ of the blob, I find
\beq
{\delta x\over w(t)}\propto {t^{3/2-d/4}\over t^{1/2}}\propto t^{1-d/4}
\label{delx/w}
\eeq
Note that this ratio diverges as $t\to\infty$ for $d<4$. Therefore, $d=4$ is another critical dimension, below which the assumption that information transport is dominated by diffusion breaks down. This is a "breakdown of linearized hydrodynamics": the linear theory, which ignores the convective $\lambda_1\bvp\cdot{\bf\nabla}\bvp$ term - that is, the term that contains the physics of the lateral wandering of the ``blob" just calculated- is incorrect: that non-linear term-i.e., the convective wandering of the blob- actually dominates over the linear diffusive process.

What happens for $d<4$? Well, now the blob must become anisotropic, since information is transmitted much more rapidly perpendicular to the mean direction of flock motion, while parallel to the mean direction of flock motion, it's still diffusive, since the speed of the flockers does not fluctuate much (it isn't a Goldstone mode), and it doesn't vary much as the direction $\theta$ fluctuates ($\vpa\sim\cO(\theta^2)$, versus $\bvp\sim\cO(\theta)$). So it looks like figure (\ref{blobright}), 
\begin{figure}
 \includegraphics[width=1.0\textwidth]{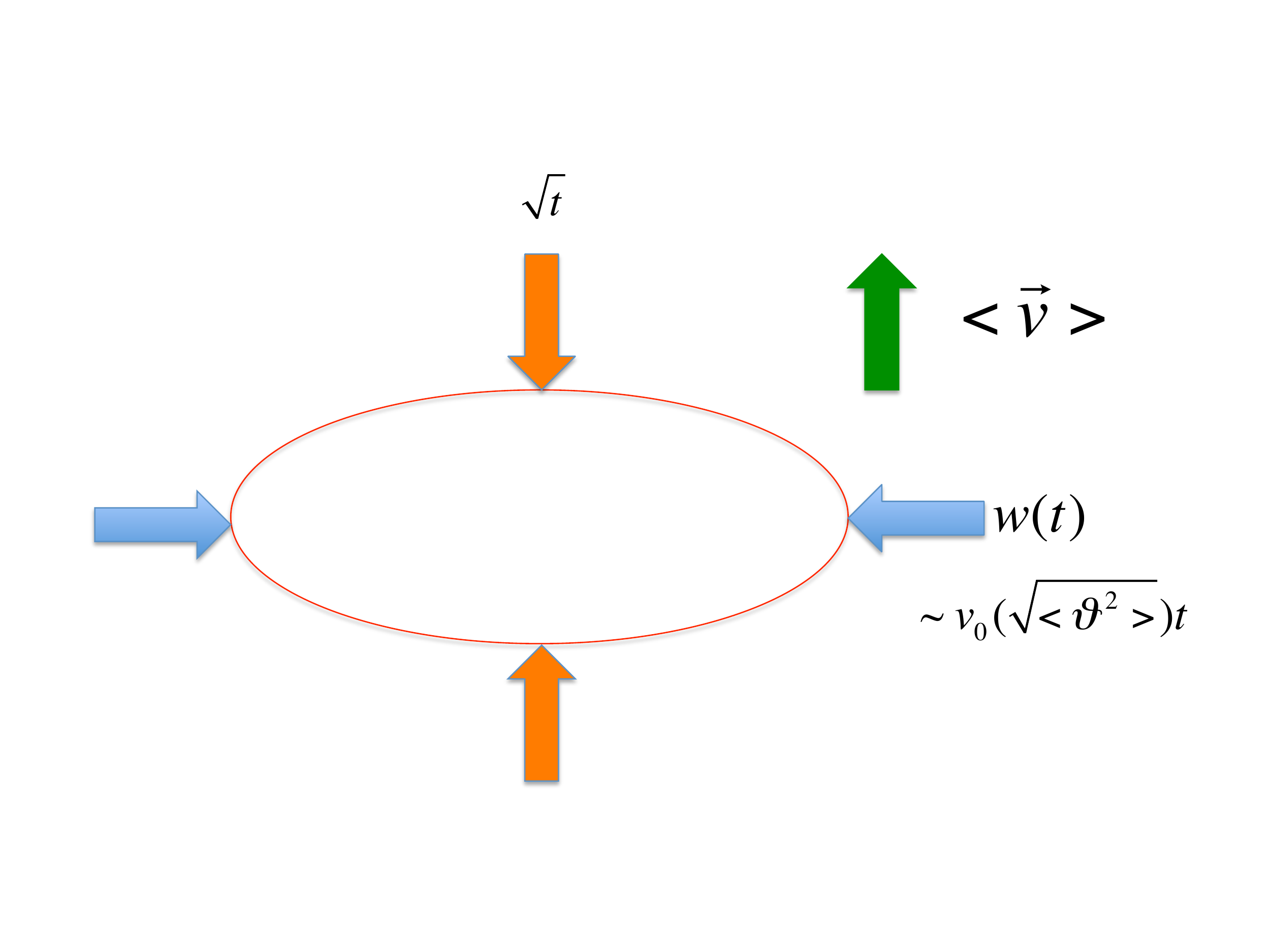}
\caption{\label{blobright}Correct picture of the evolving blob. It still  grows diffusively in the direction of the mean flock velocity (vertical in this figure), but grows laterally at a rate dominated by motion of the flockers (i.e., convection), rather than diffusion. This convective width $w(t)$ can be determined self-consistently via equation (\ref{wsc}). }
\end{figure}
where the length of the blob  along the mean direction of flock motion still grows diffusively (i.e., like $\sqrt{t}$), while the spatial extent $w(t)$ of the blob in all of the $d-1$ directions perpendicular to $<\bv>$ is controlled by the fluctuations of $\bvp$, which I will have to determine self-consistently. I'll do so by the same sort of sloppy handwaving argument I used in section II to "derive" the Mermin-Wagner theorem. That is, I'll note that
\beq
\#\,{\rm of\, errors/flocker}\propto t \,.
\label{number of errors/flocker}
\eeq
The number of flockers making errors is
\beq
N(t)\propto {\rm volume \, of \, blob}\propto  [w(t)]^{d-1} \sqrt{t} \,.
\label{Nflock}
\eeq
Hence, the total number of errors made inside the blob after a time $t$ is  
\beq
{\rm total}\, \#\, {\rm of\, errors}\propto N(t)t \,.
\label{number of errors flock}
\eeq
This gives for the rms fluctuations of $\theta$
\beq
\sqrt{<\theta^2>}\approx{\sqrt{{\rm total}\, \# \,{\rm of}\, {\rm errors}}\over N(t)}\propto {\sqrt{N(t)t}\over N(t)}\propto {\sqrt{t\over N(t)}}\propto t^{1/4}w^{(1-d)/2} \,.
\label{thetaflucflock}
\eeq
As for our earlier estimate of $\delta x$, we can estimate the lateral width $w(t)$ as
\beq
w(t)\sim v_0\sqrt{<\theta^2>}t\propto t^{5/4}w^{(1-d)/2}
\label{wsc}
\eeq
This is the promised self-consistent condition on $w(t)$. It is easily solved to give
\beq
w(t)\propto t^{5\over2(d+1)}
\label{wscale}
\eeq
which can be inverted to give the dynamical exponent $z$:
\beq
t(w)\propto w^{z} \,,
\label{tw}
\eeq
with
\beq
z={2(d+1)\over5} \,.
\label{zhand}
\eeq
Note that this is the same result I got from dynamical RG argument of section IV!
I can get the anisotropy exponent by using the fact that the blob still grows diffusively in the parallel direction:
\beq
w_{\parallel}\propto t(w)^{1/2}\propto w^\zeta \,,
\label{number of errors}
\eeq
with 
\beq
\zeta=z/2={(d+1)\over5} \,,
\label{zetahand}
\eeq
which also agrees with the result of the RG analysis of the preceding section. (To obtain the second proportionality in (\ref{zetahand}), I've used the relation (\ref{number of errors}) to relate $t(w)$ to $w$.

Finally, using (\ref{tw}) in (\ref{thetaflucflock}), I get
\beq
\sqrt{<\theta^2>}\propto w^\chi
\label{thetaflucflock2}
\eeq
with
\beq
\chi=z/4+(1-d)/2=(3-2d)/5
\label{chihand}
\eeq
which, once again, agrees with the RG result. It also predicts that $\chi=-1/5$ in $d=2$; since this is $<0$, this implies that long-ranged orientational order is stable in two-dimensional flocks.

\section{Acknowledgments}
I thank  Yuhai Tu for his years of fruitful collaboration with me on this problem, and for graciously allowing me to use some jointly written material in these notes.  I also thank T. Vicsek for introducing us to this
problem,    J.\ Sethna and K.\
Dahmen for pointing out the existence of the $\lambda_2$ and
$\lambda_3$ terms, and the organizers for inviting me to speak here at Les Houches.

\newpage


\begin{thebibliography}{000}
\bibitem{spinspace}   The isotropic Heisenberg model of magnetism
is invariant under uniform rotation of all the spins, without a corresponding
rotation of the lattice on which they live. A flock, like an
ordinary collection of interacting molecules, such as those which
form liquid crystals,
is invariant only under spatial rotations, which rotate both
the position and the velocity vectors of the creatures of the flock.
\bibitem{boidsterm} I will frequently use the term ``boid'' (a short
form for ``birdoid''), coined by C. Reynolds in the first reference
in \cite{boids}.
\bibitem{galinvfoot} This lack of Galilean invariance is appropriate if only  
 the flock is dynamical (i.e., if the background dissipative medium is static). The fully Galilean-invariant treatment of 
the coupled dynamics of a flock and a dynamical  fluid medium through which 
it moves is discussed in \cite{SR} and \cite{TT4}.
\bibitem{chirho} One could more generally rescale $\delta\rho$ with a different rescaling exponent $\chi_\rho$ from the exponent $\chi$ used for $\vec{v}_{_\perp}$. However, since fluctuations  of $\delta\rho$ and  $\vec{v}_{_\perp}$ have the same scaling with distance as time, it is most convenient to rescale them the same way. None of my conclusions would be changed if I did not make this arbitrary choice.
\bibitem{couplings}Other choices of rescaling force one to find suitable ``dimensionless couplings"; that is, ratios of the nonlinear parameters (e.g., $\lambda_1^0$) and suitable powers of the linear parameters (e.g., $\Delta$, $D_B$)  that actually give the ratios of the renormalizations of the linear parameters by fluctuations to their linear values. These necessarily have the same RG eigenvalues as those  we found here for $\lambda_1^0$,  etc. alone, independent of the arbitrary choice of $z$, $\chi$, and $\zeta$.
\bibitem{pseudo} In order for {\it both} equations of motion ($\rho$'s and $\vec{v}_{_{\perp}}$'s) to 
have this pseudo-Galilean invariance, it is actually also necessary that $w_1=\lambda_1^0$. 
However, by a more complicated argument, one can show that the existence 
of a pseudo-Galilean invariance for one value of $w_1$ ensures the same relation between exponents for all values of $w_1$ as that obtained when $w_1=\lambda^0$.
\bibitem{no log} It might appear from this argument that it would be possible for the divergence to be suppressed by only a power of the logarithm of length scale. Such a thing {\it could} happen were we in precisely the critical dimension of the problem. However, here the critical dimension is$d= $4, while the dimension under discussion here  is $d=2$. In every other problem of which I know (with the exception of problems involving quenched disorder, which are known to be a very special case), the anomalous behavior of correlation functions is algebraic in wavenumber q once one goes below the critical dimension by a finite amount. This means here that the equal time averages $< |\vec{v}_\perp(\vec{q}, t)|^2>$ must be smaller than $1/q^2$ by a power of $q$; hence, the integral of this over all $\vec{q}$, which gives the real space fluctuations $< |\vec{v}_\perp(\vec{r}, t)|^2>$, must be finite. This leads unavoidably to the conclusion that long ranged order must be stable in $d=2$.
\end{thebibliography}
\end{document}